\documentclass[journal]{IEEEtran}
\usepackage[strings]{underscore}
\usepackage{graphicx}
\usepackage{gensymb}
\usepackage{amsmath}
\usepackage{amssymb}
\usepackage{mathtools}
\newcommand{\BigO}[1]{\ensuremath{\operatorname{O}\bigl(#1\bigr)}}
\usepackage{array}
\usepackage[caption=false, ...]{subfig}
\usepackage[font=small,labelfont=bf,
   justification=justified,
   format=plain]{caption}
\usepackage{silence}
\graphicspath{ {./} }
\usepackage[normalem]{ulem}
\usepackage{color}
\usepackage{comment}
\usepackage{filecontents,lipsum}
\usepackage[noadjust]{cite}
\useunder{\uline}{\ul}{}

\ifCLASSINFOpdf
\else
\fi

\begin{document}
\title{Security Vulnerabilities Against \\ 
	Fingerprint Biometric System}
\author{Mahesh Joshi\textsuperscript{1} Bodhisatwa Mazumdar\textsuperscript{1} Somnath Dey\textsuperscript{1} \\
        \{phd1701101004, bodhisatwa, somnathd\}@iiti.ac.in \\ {\textsuperscript{1}Indian Institute of Technology Indore, India}% <-this % stops a space
		}
\maketitle

\section*{Abstract}
\textbf{The biometric system is an automatic identification and authentication system that uses unique biological traits, such as fingerprint, face, iris, voice, retina, etc. of an individual. Of all these systems, fingerprint biometric system is the most widely used because of its low cost, high matching speed, and relatively high matching accuracy. Due to the high efficiency of fingerprint biometric system in verifying a legitimate user, numerous government and private organizations are using this system for security purpose. This paper provides an overview of the fingerprint biometric system and gives details about various current security aspects related to the system. The security concerns that we address include multiple attacks on the system, associated threat models, biometric cryptosystems, current issues, challenges, opportunities, and open problems that exist in present day fingerprint biometric systems.}

\section*{Keywords}
\textbf{biometric system, biometric attacks, threat model, security, vulnerability template protection schemes}

\section{Introduction}\label{introduction}

Traditionally, passwords and tokens are in use as standard user authentication mechanisms. However, in recent times, biometric systems have ushered a new dimension to access control systems, and are widely used as a reliable and convenient means for user authentication. The primary purpose of using a biometric system is to provide non-repudiable authentication \cite{DBLP:books/daglib/p/JainNN13}. These systems have become a trusted authentication mechanism in health-care, financial, retail, education, manufacturing, military, and law enforcement agencies \cite{Maltoni:2009:HFR:1557624}. They are also used in applications for the benefit of society such as tracking child vaccination schedules, identifying missing children, and preventing fraud in food subsidies, etc \cite{DBLP:conf/ictd/JainAB0SBK16}.

Biometric systems recognize patterns, such as veins in the palm, the texture of an iris, or the minutiae of a fingerprint \cite{Jain2006ATO}. The fingerprint biometric system is a touch-based recognition system, with comparatively small finger surface exposed to the sensor for feature extraction and comparison, thus leading to relatively high matching speed and accuracy with moderate cost. Due to these advantages, it has become the most widely used and most convenient biometric authentication system. The Unique Identification Authority of India (UIDAI) initiated by Government of India has so far issued more than $1110$ million Aadhaar numbers the residents of India \cite{aadhaar}. In this process, the biometric traits such as the fingerprints are collected from every citizen to provide them with targeted delivery of financial and other subsidies, benefits, and services. The Aadhaar process presents an alert to emphasize the security aspect of the biometric system to preserve the privacy and security of biometric traits of millions of people. 

Even though  biometric systems are reliable means of authentication as compared to password or token based system, it is not completely secure and have privacy concerns to worry about. The security concerns of biometric-based application include the risks of stolen biometrics, replacing compromised biometrics, frauds done by administrators, denial of service, and intrusion etc. The biometric system also has limitation  due to which it falsely accept an impostor or falsely rejects a genuine user. The communication channel is highly insecure leading to interception of biometric data and thus losing the privacy of an individual. In this paper, we provide an insight into the fingerprint biometric recognition system and address security concerns that can be a threat to the system. We present possible attacks on the system components and communication channels between them, and ways to mitigate such attacks.

The organization of our paper is as follows. Section \ref{system} discusses the fingerprint biometric system briefly and its limitations. Section \ref{ourmodel} identifies fifteen vulnerable points in the system, where the possibilities of an attack exist. Section \ref{attacks} describes the attacks on the biometric system and their countermeasures. Section \ref{models} provides a glimpse of threat models available in the literature. Section \ref{template} provides an insight into the methods used for protecting the fingerprint template, such as cancelable biometrics and biometric cryptosystems. This section focuses on the application of cryptography in securing biometric system. In Section \ref{issues}, we present the issues and challenges for biometric recognition system. The research opportunities, and open problems in the biometric system are briefly discussed in Section \ref{problems}. Section~\ref{conclusion} draws the conclusion of the paper.

\section{Fingerprint biometric system}\label{system}

\begin{figure}[!ht]
	\centering	
	\includegraphics[width=0.50\textwidth]{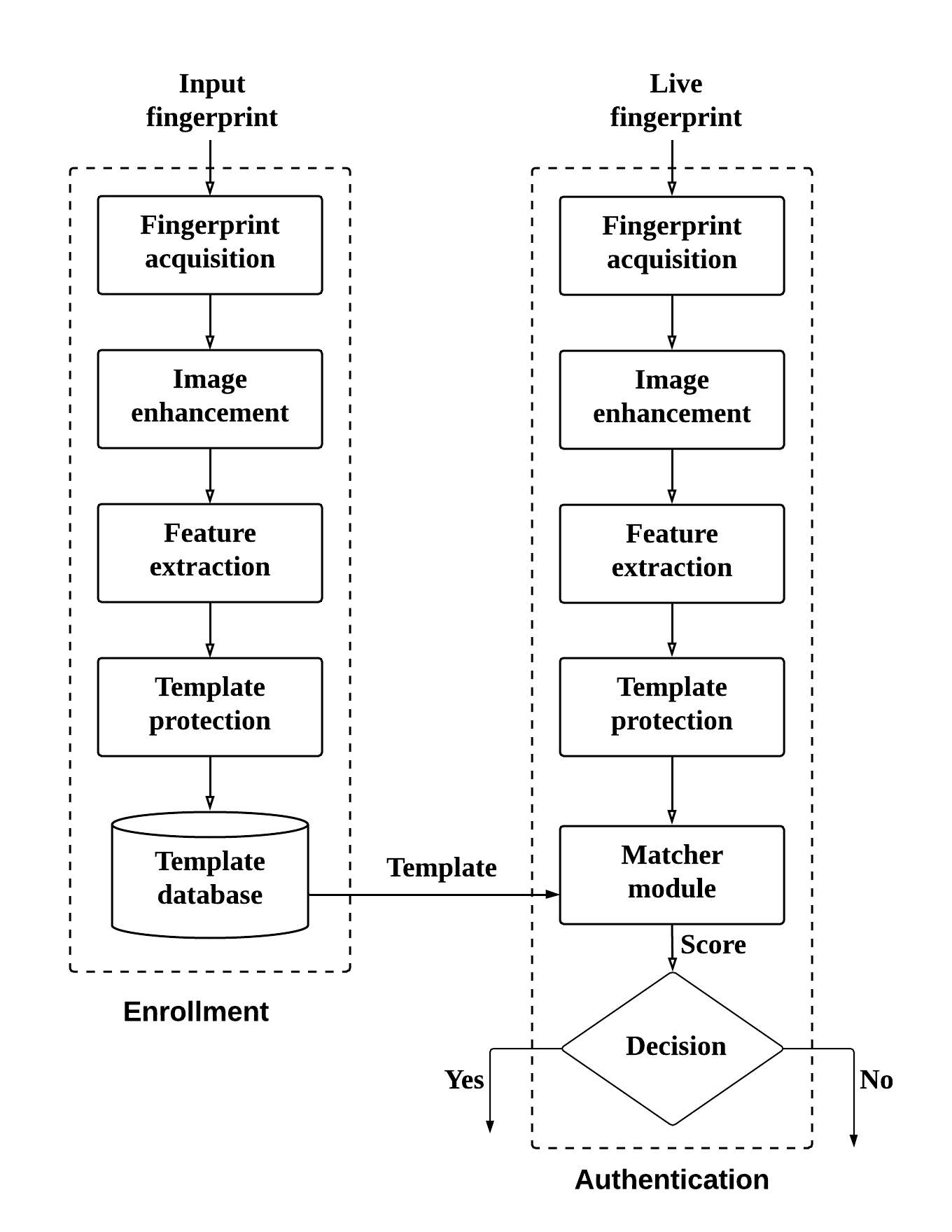}
	\caption{Fingerprint biometric system}	
	\label{fig:process}
\end{figure}

The fingerprint biometric system is a pattern recognition technique that recognizes a person based on a feature vector extracted from the person's fingerprint pattern. The steps involved in a fingerprint biometric recognition system is shown in Fig. \ref{fig:process}. These systems work in two phases, namely, an enrollment phase, and an authentication phase.

\begin{figure}[b]
	\centering	
	\includegraphics[width=0.30\textwidth]{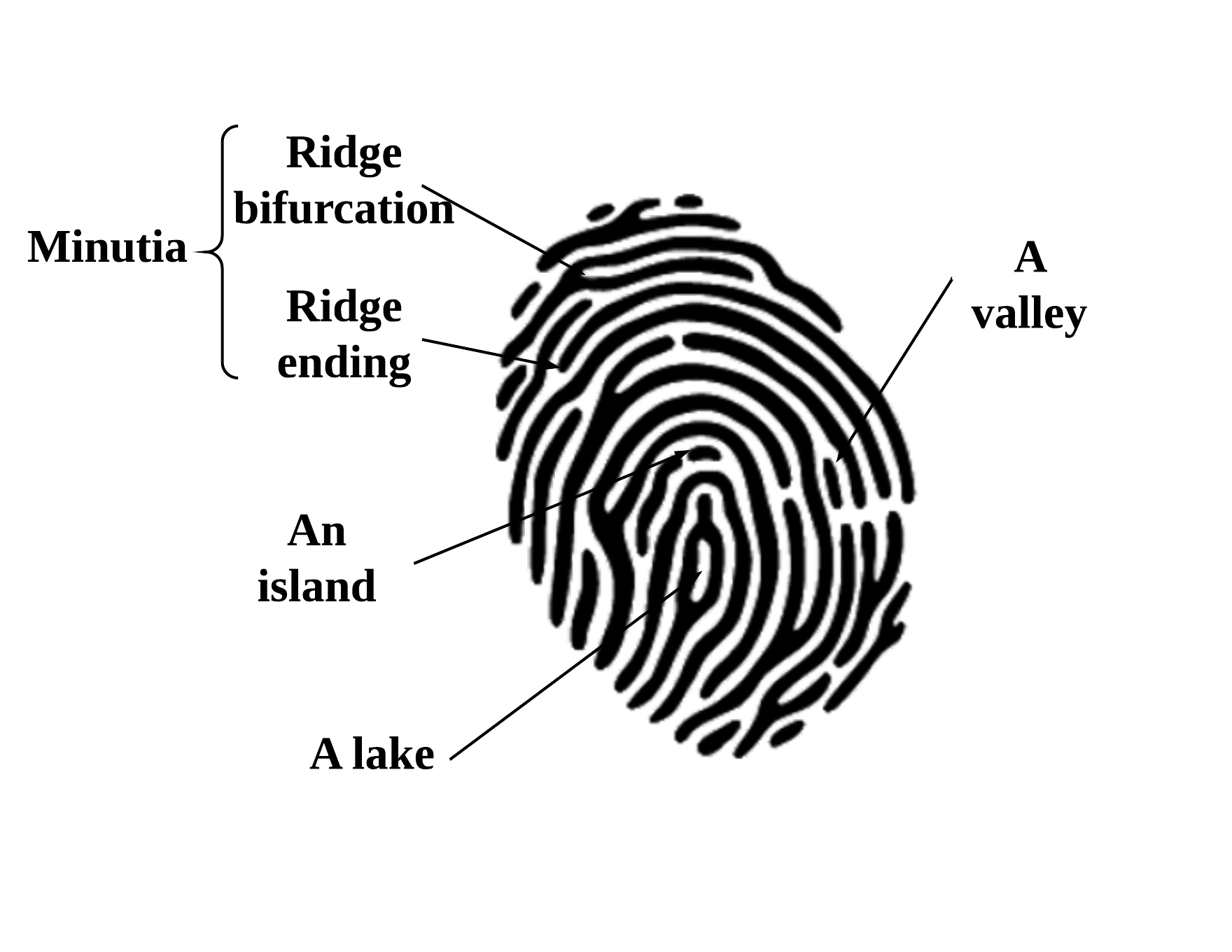}
	\caption{Features from fingerprint image}	
	\label{fig:patterns}
\end{figure}

In the \textit{enrollment phase}, the system administrator performs the registration of an individual to the biometric application. The user puts his finger on the surface of the sensor in an input device. A user of the fingerprint biometric system interacts with the sensing device as he puts his finger on the sensor. The quality of fingerprint captured by the sensor depends on the state of the finger, such as a wet or dry finger, a cut on the finger, the pressure applied by finger, the durability of the sensor, etc., and environmental factors, such as humidity,  and temperature. The technologies used for manufacturing fingerprint sensor include an optical sensor, capacitive sensor, pressure sensor, thermal sensor, an ultrasound sensor, etc. The image collected by the sensor is enhanced using image enhancement module. The feature extraction module retrieves the interesting features, shown in Fig. \ref{fig:patterns}, from the fingerprint image. The template protection module generates an encrypted biometric template that is stored in a template database. The administrator enrolls the user by assigning his credentials to the captured fingerprint.

The \textit{authentication phase} verifies and identifies the claimed individual. The ``verification" involves checking whether the presented biometric belongs to the claimed person. The ``identification" discloses who is associated with the submitted biometric. In this phase, a user puts his finger on the system and claims himself as a registered user. This attempt is termed as a live fingerprint.

The interesting features of a fingerprint image are shown in Fig. \ref{fig:patterns}. The impression where the finger surface touches the sensor forms dark lines in the image is termed as \textit{ridges}, whereas the remaining surface is called \textit{valley}. The flow pattern of ridges in a fingerprint is unique to the person in that no two people with the same fingerprints have yet been found \cite{DBLP:conf/eccv/BolleSRP02}. The ridges on the fingerprint image form different shapes, such as a lake (a closed ridge surrounding a valley), the bifurcation of a single ridge, a short ridge, a dot (or an island), bridge, crossover, etc. These complex characteristics can be represented as a combination of two basic features, i.e., \textit{ridge ending} and \textit{ridge bifurcation}. The points at ridge ending and ridge bifurcation form the \textit{minutia} of a fingerprint \cite{Maltoni:2009:HFR:1557624}.

An image quality checker follows the scanner in the system. The existing fingerprint quality metrics include, {\em segmentation-based approaches}, {\em sum of pixels standard deviation} (STD) of local block, NIST {\em fingerprint image quality} (NFIQ) level ($1-5$), etc. An NFIQ value of one  indicates high quality samples, whereas the NFIQ value of five indicates poor quality samples. A good quality captured image depicts high contrast and bright ridge structure. An image quality assessment is performed to detect a poorly captured fingerprint. Greenberg et al. \cite{GREENBERG2002227} performed experiments to compare various fingerprint image enhancement techniques and found that the techniques based on direct gray-scale enhancement perform better than approaches which require binarization (the process of converting a gray-scale image into a binary image) and thinning as intermediate steps.

The enhanced image ridges are traced to create a binary image. This binary image skeleton undergoes thinning operation to facilitate the extraction of minutia location, namely, the points of ridge ending and ridge bifurcation. A template normally contains only the information necessary for comparison in the matcher module. ISO/IEC JTC1/SC37 for standardization specifies three formats for fingerprint data as minutia-based, pattern-based, and image-based. A \textit{minutia template} contains only the minutia information, such as, position, type, and angle etc. A \textit{pattern template} contains only pattern related information such as ridge structure. An \textit{image template} typically stores the unmodified output of the sensor device, such as the image of a fingerprint. The stored template should not reveal any data that can be replayed, and it should be difficult for an adversary to guess or reverse engineer the original biometric trait or any close replica from the stored data \cite{DBLP:journals/ejasp/JainNN08}. 

A stored template satisfying the following four characteristics is termed as protected: 
\begin{enumerate}
	\item The stored template must be irreversible in generating the original biometric data. 
	\item The matching score should not vary considerably due to acquisition noise or environmental changes. 
	\item It must ensure the user's privacy. 
	\item It must guarantee that the encrypted template retrieved by the adversary from one database cannot be used for matching in another database for the same user without his consent.
\end{enumerate}
The biometric template protection schemes aim at preserving the user's privacy while enhancing the security of stored templates. Section \ref{template} presents a detailed discussion of present day template protection schemes.

\subsubsection*{Intrinsic limitations of biometric system} 

The matcher module compares the live input fingerprint template with the stored fingerprint templates. A \textit{false match error rate} or {\em false accept rate} (FAR) specifies the percentage of cases (or the probability with which) the system accepts an intruder (non-enrolled) user. It is the ratio of number of attempts made by non-enrolled users against the matching score when the system accepts some of them as authorized user. The \textit{false non-match error rate} or {\em false reject rate} (FRR) determines the percentage of cases (or the probability with which) the system rejects an genuine user. It is the ratio of number of attempts made by enrolled users against the matching score when the system rejects some of them as unauthorized user. The predefined \textit{decision threshold} value of a biometric system is set to a default value according to the requirement of a particular organization. A high security application will have a higher threshold, which means a high FRR and low FAR. A university attendance system or a customer service facility center on the other hand has a comparatively low threshold. The system's decision depends on the {\em matching score} (a numerical value) which must be greater than or equal to a predefined threshold for a given biometric system. High-performance fingerprint recognition systems can support error rates in the range of $10^{-6}$ for false accept cases and $10^{-4}$ for false reject cases\cite{Ratha}.

\begin{figure}[t]
	\centering	
	\includegraphics[width=0.50\textwidth]{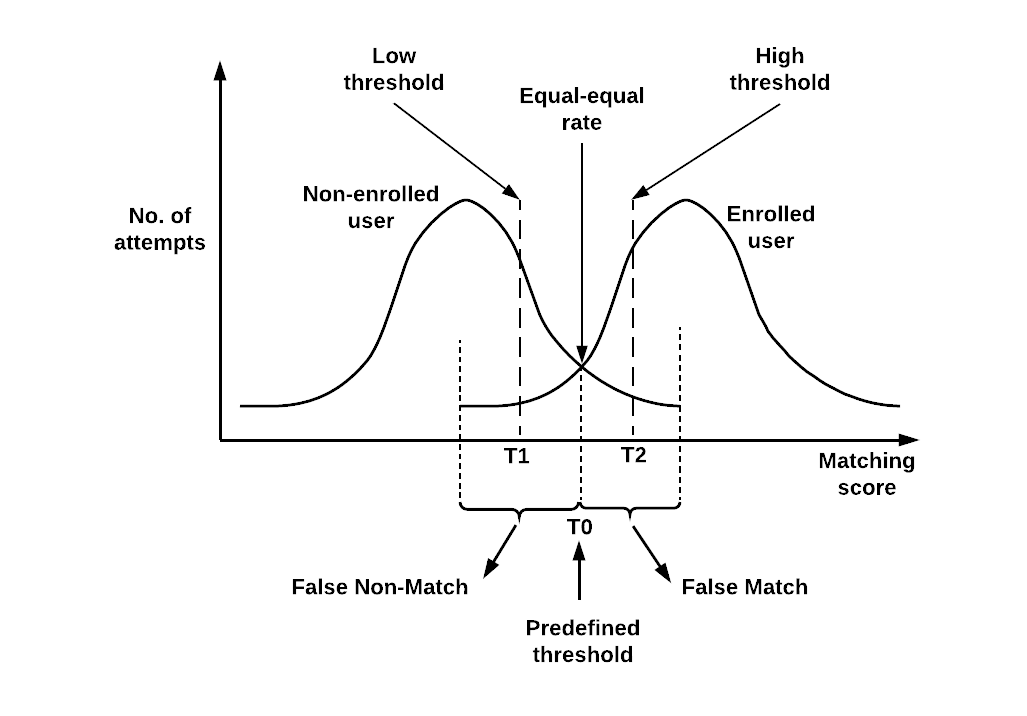}
	\caption{Error distribution curves depicting matching scores for an enrolled user and a non-enrolled user over a number matching attempts.}	
	\label{fig:error}
\end{figure} 

\begin{figure*}[!h]
	\includegraphics[width=\textwidth,height=14cm]{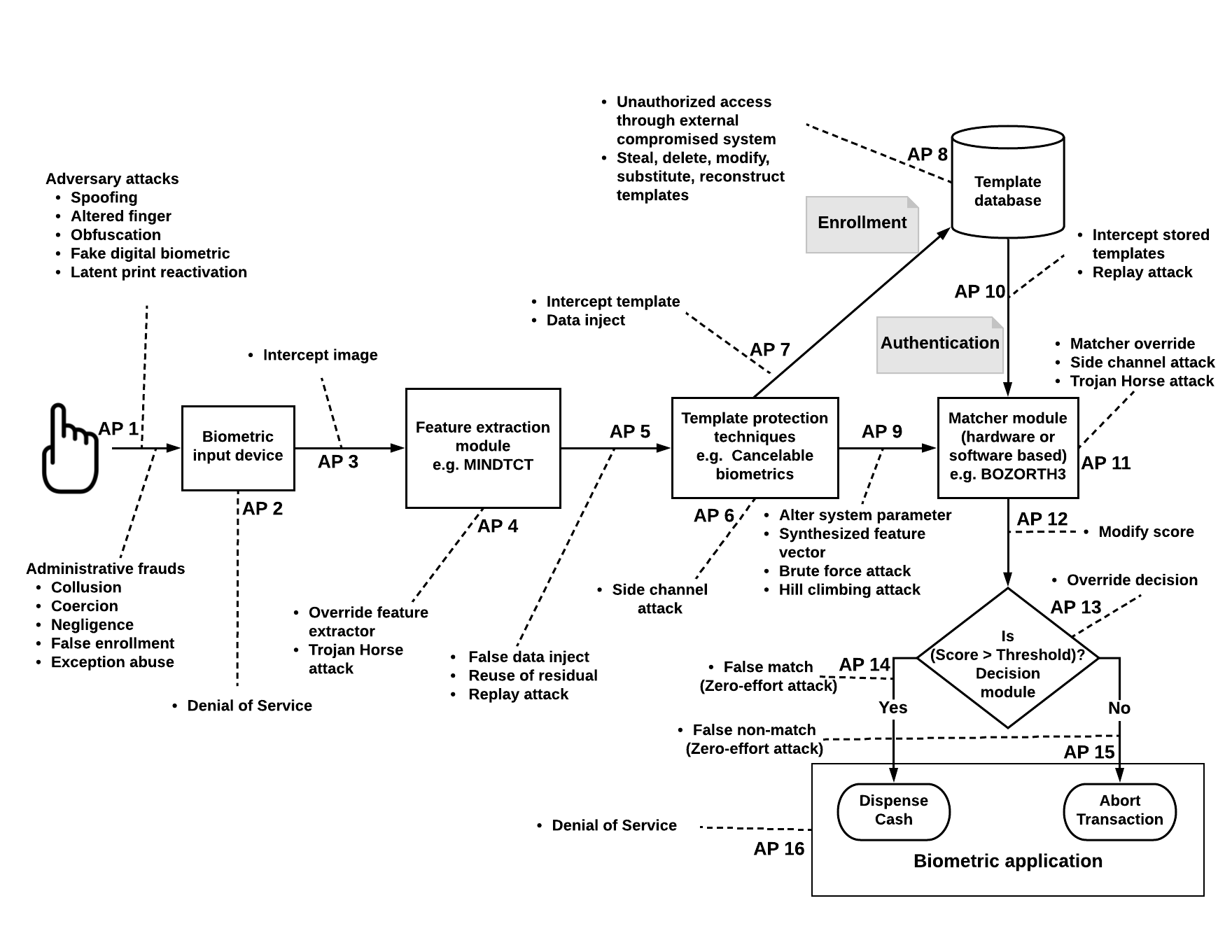}
	\caption{Vulnerabilities and attacks on different components of a fingerprint biometric system. A number corresponding to an arrow or a dashed line represents an attack point (AP).}
	\label{fig:ourmodel}
\end{figure*}

Fig. \ref{fig:error} shows the existence of a false match and a false non-match decision as a limitation of a biometric system.  For a large number of attempts by enrolled and non-enrolled users, the matches form a bell-shaped curve which overlap each other. A high threshold, \textit{T2}, removes all bad matches at the cost of allowing false-non matches. If the threshold is set to a lower value, \textit{T1}, it eliminates most of the bad matches but favors some false-matches. The \textit{equal-equal error rate} (EER) for a fixed threshold occurs when the system's false match and false non-match error rates are equal. Its value indicates the probability where the FAR and FRR are same. The equal-equal error rate is also termed as cross-over error rate. A biometric system with lower EER will be more accurate.

\section{Fingerprint Biometric system : From security perspective}\label{ourmodel}

Roberts identified the source of vulnerabilities for the biometric system that  generated out of threats. These sources are categorized into three types, namely, threat agent, threat vectors and system vulnerabilities \cite{Roberts:2007:BAV:2639537.2639843}. A threat agent is an unauthorized individual who accesses the biometric system. Thus, an attacker, an impostor, or even a legitimate user can be a possible threat to the biometric system. All possible attacks on the biometric system combine to form the threat vector.

We identify $16$ \textit{attack points} (AP) on the fingerprint biometric system as shown in Fig. \ref{fig:ourmodel}. The system components and the insecure communication channel between these components form the probable attack points that render the system susceptible to numerous attacks. The possibility of attacks is mainly attributed to the design flaws in the system. The system components in Fig. \ref{fig:ourmodel} are depicted in closed boxes. The arrow pointing from one component to another shows the transmission of the output of one component to be fed as input to the next component through a communication channel. 

The security model in Fig. \ref{fig:ourmodel} is unique in several aspects as compared to the existing threat models. We include the template protection techniques' module and identify the presence of an attack on it that aim to reveal the biometric key for the cryptosystem. The successful attempts of {\em side-channel attacks} targeting the circuit technology implementation of the biometric system are also included in our security model. During enrollment phase, the adversary can intercept the secured template over the channel connecting the template protection technique module and the database. We address different attacks on the $16$ attack points in the biometric system in Section \ref{attacks}.   

\section{Attack scenarios}\label{attacks}
The fingerprint biometric system can be the target of the system administrator, an authorized user, a layman mounting a DoS attack on the biometric input instrument, or the biometric application such as an automated teller machine, or an adversary possessing a knowledge of the biometric system. The scenarios involving the system administrator are termed as \textit{insider attacks} or \textit{administrative frauds}. 

\subsection{Direct and Indirect Attack} The biometric attacks are broadly classified as \textit{direct attacks} (also known as presentation attacks) and \textit{indirect attacks} \cite{DBLP:series/acvpr/978-1-4471-6523-1}. The adversary executes \textit{direct attacks} by presenting the biometric characteristics of a registered user to the sensing device to access the system as authorized user. The attacker targets the input device or the application controlled by the biometric system for direct attack by damaging them, rendering the system inaccessible to such users. In Fig.~\ref{fig:ourmodel} these attacks are depicted as attack points, AP 1, AP 2, and AP 16. As the sensing device is an external component of the system and is accessible to everyone, an attacker with no knowledge of the internals of the biometric system can mount this attack \cite{DBLP:journals/ieeesp/AkhtarMF15}. 

Galbally et al. performed vulnerability evaluation of fingerprint verification systems against direct attack that was carried out with fake fingertips created from minutia templates in \cite{DBLP:conf/icpr/GalballyCLMF08}. Further, the authors conducted direct attack experiments using a highly competitive ISO minutia-based matcher and fake fingers generated from ISO/IEC 19794-2 templates printed on plastic cards as 2D barcodes in \cite{DBLP:journals/prl/GalballyCLRMFOM10}. The authors used the FVC2006 DB2 database (FVC, 2006), captured with the optical Biometrika FX3000 sensor. The work showed that over $75$\% of direct attack attempts granted access to the system. Matsumoto et al. demonstrated that $11$ different fingerprint recognition systems accepted artificially created gummy (gelatin) fingers \cite{DBLP:journals/dud/MatsumotoMYH02}.

The \textit{indirect attacks} expect the attacker to have an expertise in biometric systems. In order to mount an indirect attack, the adversary must know about the internals of the targeted system component. The person directly involved in these types of attack can be an impostor or an authorized user. The interception of information transmitted over the communication channel, which target the internal components of the biometric system to override its output or manipulate the stored template are examples of indirect attacks. The attack points other than AP $1$, AP $2$, and AP $16$ in Fig. \ref{fig:ourmodel} fall under this category.

\subsection{Repudiation}
An individual can take the benefit of the intrinsic limitation of the biometric system. The attacker can refuse to have accessed the system by arguing the false acceptance phenomenon associated with the system. The bank technician who withdrew cash using the biometric application specified in Fig. \ref{fig:ourmodel} can point out the possibility of false acceptance by the biometric system. 

\subsection{Contamination or covert acquisition}
The authorized user can unknowingly leave behind the traces of his fingerprints in daily routine. The adversary uses the fingerprint traces to generate fake fingerprint using spoofing or constructing a mold by lifting latent fingerprint to mount the attack. An attempt to regenerate fingerprint image from a stolen template is also an example of such attack. The attacker can acquire the biometric of a user from biometric home security application and use it to dispense cash from the biometric application in Fig. \ref{fig:ourmodel}. 

\subsection{Coercion}
The attacker can forcefully ask the authorized user to submit his biometrics to the input device, for example, at gunpoint. The attacker uses the biometric application with the user's biometric data for financial transactions. The possibility of such incidents is high at automated teller machines having an insecure infrastructure, no security personnel, improper security measures like CCTV cameras. 

\subsection{Negligence}
A legitimate user can hurriedly forget to log out from the biometric application while the attacker is noticing him. The adversary exploits such an incidence of negligence. He continues the session pretending as an authorized user and can carry out some more transactions or even access sensitive information about the user

\subsection{Insider attacks}
The system administrator or the authorized person can help in mounting an insider attack on the biometric system. The administrator reveals the flaws in the system to the attacker, or the legitimate user cooperates in executing such attacks. 

\subsubsection{Collusion}
In collusion scenario, a legitimate user, such as the system administrator with full access privileges is the attacker. He illegally accesses and modifies the system parameters, such as the predefined threshold. The attacker can change the access rights of an authorized user. 

\subsubsection{Enrollment fraud or false enrollment}
The administrator is the authority involved in the enrollment phase. He can help the attacker by enrolling as a legitimate user. The administrator is satisfied with a massive bribe for the illegal enrollment for entry into the government premises with high security.

\subsubsection{Exception abuse}
The exception handling procedure designed to facilitate authorized users in case of emergency can become a vulnerability to the system. The system administrator uses this facility to help the attacker access the system as an enrolled user. The system administrator resets the predefined threshold to a lower value such that the adversary can take benefit of the false match error and become authorized.

\subsubsection{Function creep}
The administrator can collect biometric samples of an individual whose count may exceed than that needed to be stored for matching in the authentication phase. The other traits along with the personal information and government identifiers of the user like Social Security Number in US or Aadhaar number in India, can be sold to private organizations for financial or other malicious reasons. 

The system should have multiple system administrators with hierarchy having different levels of privileges so that an individual authority cannot mount an insider attack, such as collusion and exception abuse. An organization can distribute the responsibility of enrollment to several departments to avoid the possibility of an enrollment fraud. One department can issue a smart-card cum identity card to every employee whose document based verification is performed by another department. This card can be used by a third department (performing enrollment process) to store his biometrics on the card itself. 

\subsection{Sensor attacks}
The sensor attacks define how an attacker can grant access to the system with manipulated biometric submission to the sensor. The attacks on system input are specified as Type 1 in Fig. \ref{fig:ourmodel}. Since these attacks are carried out at the entry point of the biometric system, the security mechanisms implemented for digital protection, such as cryptosystems are ineffective in such scenarios.

\subsubsection{Spoofing attack}
The stolen or fake fingerprint submission to a biometric system is termed as a spoofing attack. A spoofing attack represents the presentation of false data, or a false biometric claiming to be legitimate, in an attempt to circumvent the biometric system controls \cite{Roberts:2007:BAV:2639537.2639843}. The spoofing is probably the easiest way for an attacker to become authorized. To succeed, the attacker need not know about the internals of the matching or the encryption algorithm. 

\begin{figure}[t]
	\centering	
	\includegraphics[width=0.5\textwidth]{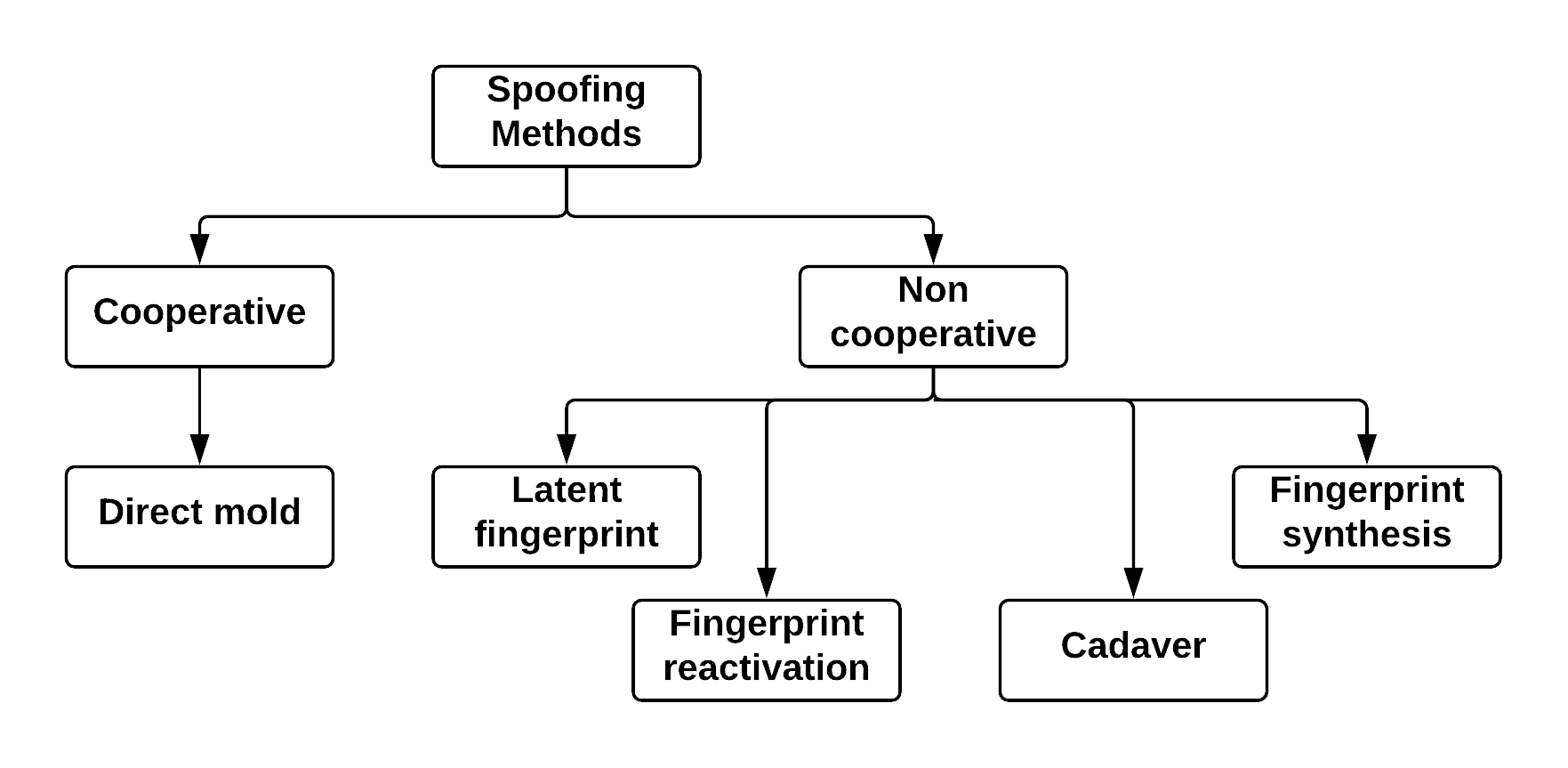}
	\caption{Methods employed for creating artificial fingerprints \cite{DBLP:journals/csur/MarascoR14}}	
	\label{fig:spoofing}
\end{figure}

Fig. \ref{fig:spoofing} shows various methods that can be used for creating artificial fingerprints. In a cooperative method, the owner of the fingerprint is involved in the process of creation of an artificial impression of his fingerprint. The direct mold of a legitimate user is created using a dental impression material or plaster. The non-cooperative methods typically create artificial fingerprints by collecting the impressions left by the user at different places. A powder and brush are used to create latent fingerprints by placing it on transparent material, such as glass. Simple techniques, such as breathing, placing a water-filled plastic bag, or brushing graphite powder on the sensor have been used to reactivate latent fingerprints deposited on a sensor \cite{DBLP:journals/csur/MarascoR14}. The use of a dead person's finger to create his artificial fingerprint is termed as {\em cadaver}. The fingerprint synthesis is a technique to recreate a fingerprint from its stored template. The fingerprint reconstruction process can be moderated by employing an iterative hill-climbing approach \cite{DBLP:journals/pami/RossSJ07}. 

\begin{figure}[b]
	\centering	
	\includegraphics[width=0.5\textwidth]{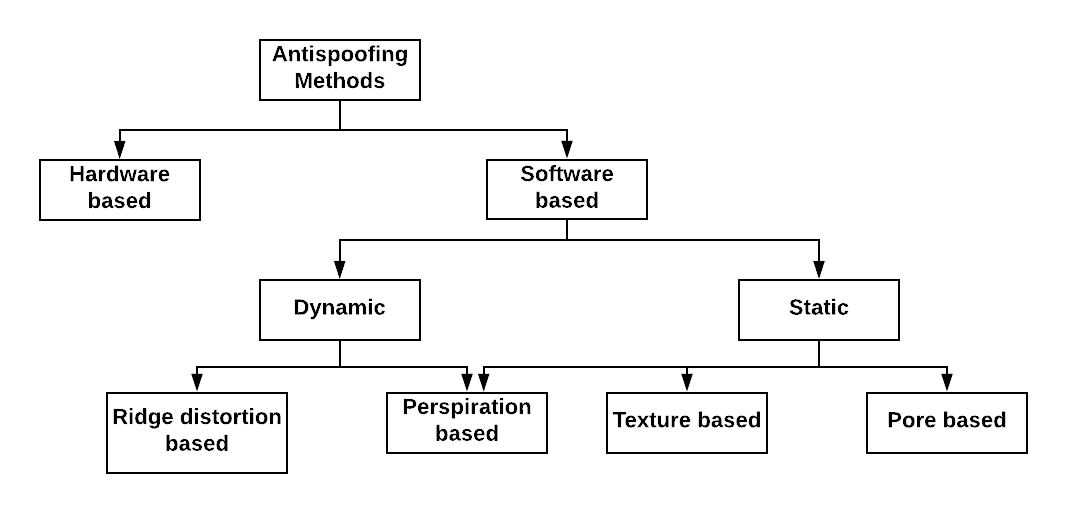}
	\caption{Existing anti-spoofing approaches \cite{DBLP:journals/csur/MarascoR14}. They are broadly classified into hardware and software based approaches.}	
	\label{fig:antispoofing}
\end{figure}

Fig. \ref{fig:antispoofing} shows various existing methods for fingerprint anti-spoofing. The hardware-based solutions are costly since a new hardware component needs to be added to the biometric sensor. Also, this solution may add a new threat to the system, so the software-based solutions were more emphasized rather than a hardware-based solution. To derive dynamic features (e.g., ridge distortion and perspiration), multiple frames of the same fingerprint image captured with two or five seconds interval are analyzed. The existence of pores on finger results in perspiration and generates an image with a darker region near pores. The spoofed finger does not possess this phenomenon, and thus generates almost same image over a short period. The change in gray level between the first and last images in a sequence can be measured by considering the local maxima and minima of the ridge signal \cite{DBLP:journals/csur/MarascoR14}. 

When a real finger is put on the sensor (multiple times in a short span), it produces distortion in the image due to pressing and moving the finger on the sensor surface. The distortion can be analyzed by observing multiple frames at high frame rate when the user presses or moves the finger. The texture properties of a finger comprises its morphology,  smoothness, and orientation. 
For a spoofed finger, these properties are different than a live finger, for instance, the live finger surface is smoother than the spoofed one. In a static pore based approach, the live and spoofed fingerprint can be detected by comparing their pore quantities. The presence of pores and locating pores on the surface is another approach used to distinguish between spoofed and live fingerprint \cite{Choi2007}.

Feng at al. proposed an algorithm to detect altered fingerprints in \cite{DBLP:conf/icpr/FengJR10}. Distorted fingerprints have unusual ridge patterns which are not found in natural fingerprints. Such fingerprints include abnormal spatial
distribution of singular points or abrupt changes in
orientation field along the scars \cite{DBLP:journals/pami/YoonFJ12}. The proposed algorithm classifies the fingerprint as a natural or altered fingerprint using a support vector machine (SVM) by employing the features extracted from ridge orientation field. It is observed that the continuous orientation field of a natural fingerprint is indeed continuous (i.e. no singularity) but for an altered fingerprint, the continuous component of the orientation field is discontinuous. The proposed algorithm extracts a feature vector, called as {\em curvature histogram}, from the continuous orientation field. The curvature histogram is input to the support vector classifier  to distinguish between a natural and an altered fingerprint. The work used real (human) and altered fingerprint in their experiments, and their proposed algorithm could detect $92$\% of altered fingerprints. 

\subsubsection{Masquerade}
A masquerade attack is a general term given to any attempt made by the adversary pretending as a legitimate user to access either the system or the information and services of a registered user. The attacker in such cases creates fake fingers, or alters the fingerprint of the authorized user to mount a spoofing attack. The anti-spoofing approaches discussed above can help in mitigating masquerade attack.

\subsubsection{Altered fingerprint submission}
Altered fingerprints are real fingers used to conceal one’s identity to evade identification by a biometric system, e.g., using an abrasive material to manipulate the ridges of one’s finger. Yoon et al. have compiled the case studies of altered fingerprints by an individual \cite{DBLP:journals/pami/YoonFJ12}. The authors classified the altered fingerprints as obliteration, distortion, and imitation. An obliterate fingerprint is a result of a cut, burn, skin disease, or side effects of drugs.

\subsubsection{Obfuscation}
Obfuscation can be defined as a deliberate attempt by an individual to mask his identity from a biometric system by altering the biometric trait prior to its acquisition by the system e.g. mutilating the ridges of one’s fingerprint by using abrasive material \cite{DBLP:journals/pami/YoonFJ12}. The altered fingerprint submission falls into the category of obfuscation. It is not guaranteed that an altered fingerprint will always succeed in evading the biometric system since the number of minutia points extracted from unaltered area before and after the mutilation are not sufficient for a successful match in every case. 

\subsubsection{Fake digital biometric}
The digitized latent fingerprints can be used by the adversary to mount a masquerade attack. The replay of a feature set to the matcher module is an example of fake digital biometric submission attack. The adversary can reconstruct intercepted template data to get authorized as multiple genuine users and thus hide his own identity.

\subsubsection{Fake physical biometric}
The spoofing is the process of generating fake physical finger of a legitimate user to enter the system. The attacker fixes his target before executing the attack to increase the chances of a successful attack. InterGov \cite{intergov} reports that insiders commit about 80 percent of all cyber crimes (an assessment based only on reported security breaches) \cite{DBLP:journals/ieeesp/PrabhakarPJ03}.  

\subsubsection{Latent print reactivation}
A latent print of the fingerprint will be generated on the biometric sensor surface due to oil secretion from sweat glands in the palm, or contact with the oily or sticky material. The impressions of the fingerprint on the sensor surface can be used by the adversary to create readable prints. In such cases, the attacker uses the fumes from cyanoacrylate glue or powder for reactivating the latent fingerprint. 

\subsection{Feature extractor attacks}
The feature extraction module is responsible for identifying minutia points from the fingerprint image generated by the input device. The adversary can target the feature extraction module such that it becomes incapable of generating the real feature corresponding to the submitted fingerprint.

\subsubsection{Override feature extractor}
The attack on feature extraction module by the adversary involves replacing the feature set extracted by the module with the feature set chosen by the attacker. This is usually conducted through an attack on the software or firmware of the biometric system \cite{Roberts:2007:BAV:2639537.2639843}. 

\subsubsection{Trojan horse attack}
\begin{figure}[h]
	\centering	
	\includegraphics[width=0.45\textwidth]{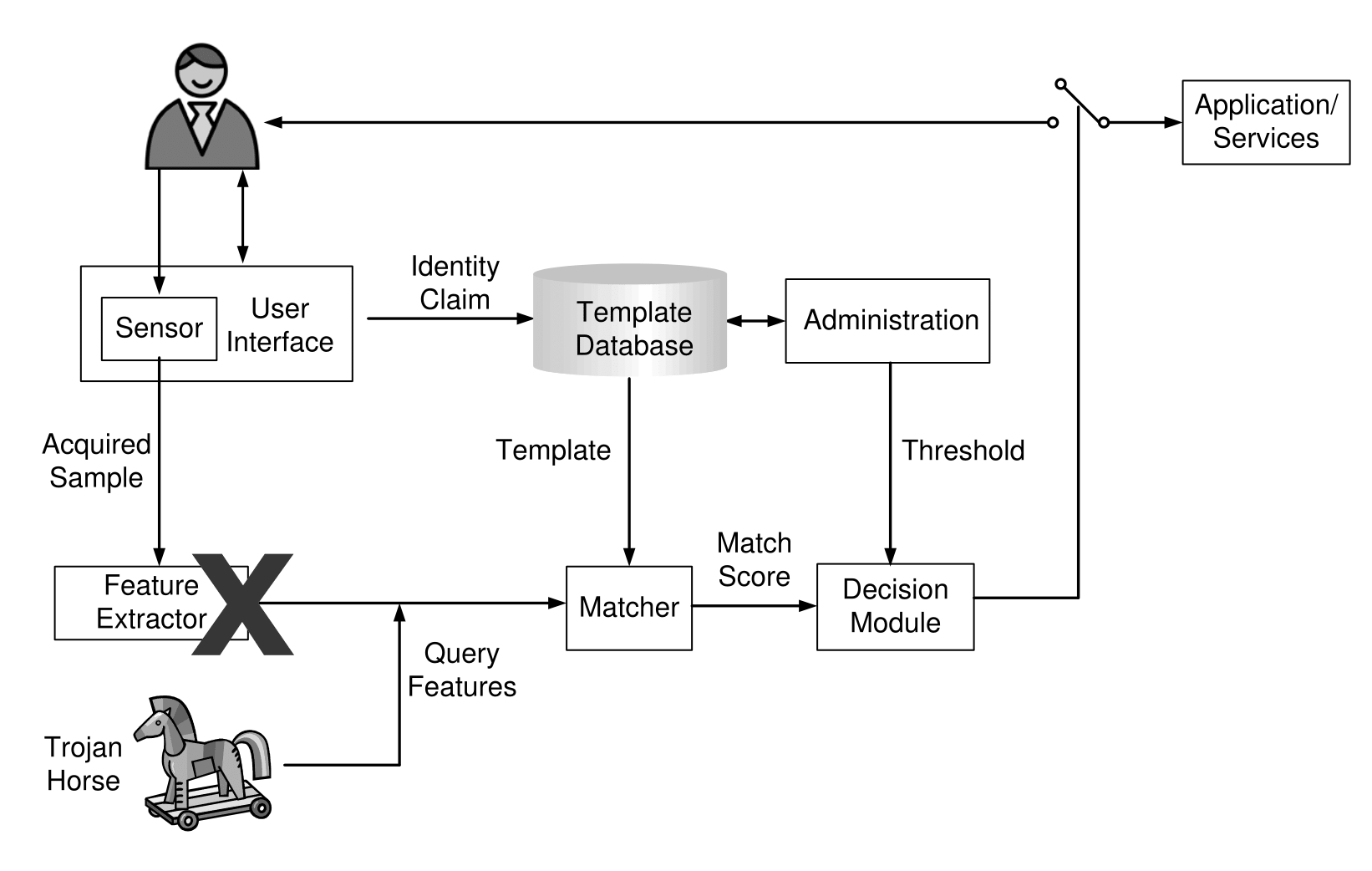}
	\caption{A Trojan Horse attack against the feature extraction module \cite[Chapter~7]{Jain:2011:IB:2161587}}
	\label{fig:trojanattack}
\end{figure}

Fig. \ref{fig:trojanattack} shows a Trojan Horse attack against the feature extraction module. It can be seen that a Trojan attack breaks the legal process of biometric system to override the feature extraction module. Trojan horse is a malicious program that can be controlled remotely by the adversary through commands. Once activated, the Trojan can delete, copy, or modify the data from the targeted system component. For such incidence in Fig. \ref{fig:ourmodel}, the Trojan will generate preprogrammed feature set to be fed as input to the template protection techniques module. 

\subsection{Attacks on template protection techniques module}
The features extracted by the feature extraction module are used to generate a secure digital template using cryptographic techniques, such as, cancelable biometrics and biometric cryptosystem, as mentioned in Section \ref{template}. The template protection module responsible for performing the translation of a feature vector into a cryptographically secured biometric template is a potential target of the attacker. The attacker aims to extract information related to the key used in the encryption algorithm. The attacks on template protection techniques module are specified as AP 6 in Fig. \ref{fig:ourmodel}.

\textit{Side channel attacks (SCA)} target a specific implementation of a circuit on a technology platform that recover the embedded secret key of the cryptosystem \cite{DBLP:books/sp/10/Standaert10}. In these attacks, the adversary uses the information leakage, such as power consumption statistics, timing information, or the electromagnetic radiation as a side channel to recover the key of the encryption algorithm. An electromagnetic field is generated due to the movement of electric charges. Electromagnetic side-channel attacks exploit correlations between intermediate data (function of the secret key) and variations in electromagnetic emanations from tamper-resistant devices, such as smart-cards \cite{DBLP:conf/ches/GandolfiMO01}. In timing side-channel attacks, the execution time for a cryptographic algorithm reveals the valuable information related to the secret code involved in the computation such as an encryption key.

\subsection{Matcher module attacks}
The matcher module in the biometric system is a hardware or software component responsible for performing the matching between the template generated from live fingerprint and the stored template during the authentication phase. The output of a matcher module is the similarity score between the two templates under comparison. The adversary targeting the matcher module aims to achieve a matching score above the predefined threshold for a non-enrolled fingerprint. The attacks on matcher module are specified as AP 11 in Fig. \ref{fig:ourmodel}. A software-based matcher executable can be digitally signed to mitigate attack by an adversary trying to execute other malicious code to override the output of matcher.

\subsubsection{Hill climbing attack}
The term hill-climbing designates an attack in which the similarity score given by the matcher is used to iteratively modify a synthetically generated template, or a group of templates, until the verification threshold is reached \cite{5507530}. The goal of the hill-climbing attack is to get authorized due to intrinsic limitation of false acceptance of the biometric system. The set up for hill climbing attack includes a database of artificially generated minutia templates (i.e. fake digital biometric) and an application that submits these templates to the matcher of the targeted biometric system. The application collects the matching score of a submitted template and modifies the template so that the matching score ultimately exceeds the decision threshold to access the system. A time-out lock-out policy that allows a specific number of unsuccessful attempts and lock the biometric system for an random time period can be used to mitigate hill-climbing attack. 

\subsubsection{Matcher override or false match}
The matcher module compares the live fingerprint template with the stored templates and preserves the score corresponding to each matching pair. The claiming user is authenticated with the identity of the enrolled user whose stored template match against the live fingerprint resulted in highest score. The adversary targeting the matcher module overrides the score generated for the genuine template comparison with a matching score above the predefined threshold. A matching score greater than the threshold dispenses cash from the biometric application shown in Fig. \ref{fig:ourmodel}.

\subsubsection{Side channel attacks}
The timing analysis involves measuring the time required for executing private key operations. Galbally et al. presented a time analysis of a reference minutia-based fingerprint matching system and studied the relationship between the score generated by the minutiae-based NIST Fingerprint Image Software 2 (NFIS2) and the time required to produce the matching score \cite{DBLP:conf/cost/GalballyCFO09}. The experimental result shows a clear correlation between the score and time required to produce the score, i.e. on average the higher a matching score the larger is the time. Furthermore, the potential side-channel leakages of a hardware fingerprint biometric comparison module was studied in \cite{DBLP:journals/iacr/BerthierBBCCDFG14}. The authors concluded that the simple power consumption analysis during matching process reveals sensitive information i.e. we can get the number of minutia for each angle value by increasing the angle value of the input minutia \cite{DBLP:journals/iacr/BerthierBBCCDFG14}.

\textit{Simple power analysis} (SPA) : The current flowing from the power supply unit to the matcher module of a biometric system can be analyzed to detect power consumption patterns that leaks the information related to the cryptographic operations. The sequence of executed microprocessor instructions can be a useful information revealed by SCA. The adversary is assumed to have knowledge about the time at which the power consumption is correlated with the secret key. The most common defenses against SCA include adding noise, dummy instructions, and power consumption balancing. 

\textit{Correlation power analysis} (CPA) : CPA is a statistical tool that uses Hamming distance for evaluation of cryptographic module leakage and/or retrieve the secret data like encryption key or minutia information in case of fingerprint biometric system. In biometric system, it is infeasible to directly link the output (matching core or the binary decision) with the input (fingerprint). So, only a CPA that considers (input data, hypothesis) pairs during the enrollment/registration phase is a feasible way to find out the correlation between the input and output. The vulnerability analysis of an embedded hardware verification module (matcher used in Match-on-card systems) against SCA by exploiting the activity leakage of the matcher to retrieve the stored trait was presented in \cite{DBLP:conf/isca/ChoutaGDBBBFC14}. The authors investigated the SCA vulnerability of the registration phase and demonstrated how side-channel analysis by CPA to retrieve a reference fingerprint is feasible on a biometric matcher module in \cite{DBLP:conf/isca/ChoutaGDBBBFC14}. They retrieved a part of the reference minutia set i.e. one reference minutia per angle, in their experiment.

\textit{Differential power analysis} (DPA) :
Applying techniques of differential power analysis to fingerprint matching is challenging due to the relatively large candidate space and the linearity of matching algorithms \cite{DBLP:conf/ccs/DurmuthOP16}. The possibility of side-channel attacks on fingerprint matching algorithms such as, the Bozorth3 and a custom matching algorithm, was studied in \cite{DBLP:conf/ccs/DurmuthOP16}. In \cite{Tiri2005ASL} the authors have presented a secure co-processor that does not leak information through the power supply. To the best of their knowledge, the authors reported that this is the first IC that is practically immune to DPA attacks. 

\textit{SCA countermeasures} : Viable cryptosystem designs must address power analysis attacks. Masking is a commonly proposed technique for defending against these side-channel attacks \cite{DBLP:conf/ches/WaddleW04}. Integrated Circuits with active shield can be used to avoid side channel leakages. Another way to mitigate side channel attacks is to modify the implementation of the cryptographic algorithm such that all computations consume same power and use equal number of clock cycles. A robust secure fingerprint matching technique, which is secure against side channel attacks was proposed by Yang et al. \cite{Yang:2003:SFM:982507.982524}. 

\subsubsection{Trojan horse attack}
The Trojan horse is a malicious executable program that suppresses the actual matching score output of the matcher module with the predefined score such that the adversary can bypass all the earlier modules (stages) in biometric system and still get authorized to access the application. This is a tampering attack where the adversary alters the vendor's matching algorithm code and replaces or modifies it with the Trojan horse code. Trojan programs can be remotely controlled  by the adversary to send commands to the matcher module.

\subsection{Template database attacks}
The template database stores encrypted biometrics and details of each enrolled user. The biometric system uses a stored template and information corresponding to it to perform the identification and verification of a user in the authentication phase. The adversary targets the database either directly or indirectly via an externally compromised system (if the server is online). The primary intention of the attacker is to reuse or tamper the stored templates. Stolen biometrics can be used to attempt a man-in-the-middle attack, such as a replay attack. The attacker can delete the stored templates to mount a DoS attack. Fig. \ref{fig:ourmodel} represents an attack on the database as AP 8. Li et al. have proposed a privacy protection scheme (hiding scheme) for the thinned fingerprint template in \cite{DBLP:journals/spl/LiK11}. In this scheme, the user's identity is hidden into thinned fingerprint image during enrollment and such a fingerprint template is stored in an online database for authentication. In the proposed data hiding scheme, no boundary pixel is produced in data embedding, resulting in a marked-thinned fingerprint of visually imperceptible abnormality without compromising the performance of the fingerprint identification \cite{DBLP:journals/spl/LiK11}. The authors claim that using such scheme, it is infeasible for the attacker to reveal the identity of the user from stolen templates collected from a compromised online template database.

\subsubsection{Template reconstruction}
The attacker breaks into the online database where the encrypted biometrics are stored by cracking a database account with administrative privileges or by exploiting the vulnerability in the database software. He modifies the existing encrypted biometrics or reconstructs a new template from them. The modified and reconstructed templates form the base for a replay attack.  If the attacker can tamper and modify the template information and generate a false ID to an illegitimate person, it is termed as a {\em template modification attack}. The adversary can use the modified biometrics to mount a {\em hill-climbing attack} at AP 9 in Fig. \ref{fig:ourmodel}. Ross et al. demonstrated that the fingerprint reconstruction process can be moderated by employing an iterative hill-climbing approach \cite{DBLP:journals/pami/RossSJ07}. The authors demonstrated that the information about orientation field, fingerprint class information namely, $A$, $L$, $R$ and $W$, and the friction ridge structure can be generated out of the minutiae template alone. A possible way to mitigate the reconstruction of fingerprint from stored template is to create  templates from encrypted minutia instead of storing minutia on template.

\subsubsection{Unauthorized template modification}
The attacker can target the online database from an externally compromised, hacked system through a known or unknown bug in the database software, log information on the disk, or information in cache or primary memory. Such an attempt is difficult to trace since the compromised system is mostly a hacked computer under the control of the attacker. The adversary can steal the  minutia templates to execute a replay attack. He can even delete and substitute the minutia templates to mount a DoS attack for the genuine users.

\subsubsection{Circumvention}
The adversary controlling the application secured with the biometric system is said to be attempting circumvention. The circumvention threat can be cast as a privacy attack, where the attacker obtains the unauthorized data (for example, accessing the medical records of another user) or, as a subversive attack, where the attacker manipulates the system (for example, changing those records, submitting bogus insurance claims, etc.) \cite{DBLP:conf/sswmc/UludagJ04}. The adversary gets himself authorized by attempting any attack, such as spoofing, replay, administrative frauds etc., to access the application and then mounts circumvention on the application data. 

\subsection{Man-in-the-middle (MiTM) attack}
In a typical scenario of a man-in-the-middle attack on the biometric system, the attacker intercepts the communication channel to collect the biometric data of a legitimate user. The attacker, in this case, can target AP 3, AP 5, AP 7, AP 9, AP 10, and AP 12 from Fig. \ref{fig:ourmodel} on the biometric system. These attacks are similar to replay attacks, where the attacker listens to the communication channel for capturing biometric data and then alters this data to access the system as a legitimate user. 

\subsubsection{Replay attack}
In the replay attack (also known as false data injection attack), an attacker bypasses the scanner and uses the previously generated digital image of a legitimate user to access the system. The attacker injects the false intercepted data directly into the communication channel to the matching component. Here, the attacker intercepts the communication medium between the sensor and the matching component, then alters the contents (to access the system as more than one genuine user), and then retransmits to the matcher for processing. This is a type of the man-in-the-middle attack represented as AP 5 in Fig. \ref{fig:ourmodel} which requires the attacker to have good knowledge of the system and database. In data injection attack, the biometric system and the template database are compromised. 

A challenge/response method to mitigate a replay attack considers every attempt involving all modules of the biometric system to authorize a user as a single transaction. It assumes that the sensor is intelligent enough to respond to a query by a secured transaction server. Whenever the sensor is touched with a user finger, the transaction server sends a query, e.g.  a random pixel value, to the sensor. If the image received by the feature extractor has the same pixel value as responded by the sensor then the transaction continues otherwise an attempt of bypassing the sensor and thus an attack is detected and the transaction is aborted. 

\subsubsection{Reuse of residuals}
The biometric system can hold the retrieved digital template in its primary memory for some time after performing the matching operation. The adversary can use the biometrics available in main memory to access the system as a legitimate user. The adversary replays the biometric templates to mount reuse of residual attack at AP 5 in Fig. \ref{fig:ourmodel}. The memory holding the data for matching can be reserved and flushed immediately after a claiming user is either rejected or accepted by the system to avoid reuse of residual by the adversary.

\subsubsection{System parameter override or modification}
The attacker manipulates the system parameters to widen the range of false acceptance and subsequently shorten the false reject margin. As a result of such alteration of system parameters (especially FAR/FRR parameters), the adversary creates a path to access the application in Fig. \ref{fig:ourmodel} with poor quality fingerprint images or incorrect data. The code signing technique can detect any tampering or alteration in the code, thus mitigating such attacks. Protection of software code against illegitimate modifications are discussed in \cite{DBLP:conf/isw/Oorschot03} and \cite{DBLP:conf/ccs/ChangA01}. 

\subsubsection{Synthesized feature vector}
The adversary with a knowledge of the biometric system can create a biometric template outside the system and directly inject it into the data stream of the communication channel. The hill-climbing attack can use this synthesized template at AP 9 in Fig. \ref{fig:ourmodel}. The challenge-response technique can help in detecting an attack involving a synthesized feature vector.

\subsubsection{Brute force attack}
A brute force attack on the biometric system involves submitting a large number of fingerprints to the system. Brute force attack is mounted at AP 4 in Fig. \ref{fig:ourmodel}. The attacker must maintain a real fingerprint template database of large number of entries in order to execute the brute force attack efficiently. The work of Ratha et al. in \cite{DBLP:conf/avbpa/RathaCB01} gives the relation between the matches that occurred for various attempts of brute force attack. A dictionary-based guessing attack is a sophisticated brute force attack. Here, the attacker chooses only those fingerprints that have a high probability of matching. The average number of attempts needed by a brute-force attack can be derived from the FAR of the system \cite{4105331}. A 0.001\% FAR indicates that 1 out of 100,000 brute force attacks on an average will be successful \cite{7948983}. 

\subsubsection{Storage channel intercept and data inject}
The communication channel between the encrypted biometric template database with other components, namely, the template protection techniques module and the matcher, can be the source of information for the adversary. The attacker may inject false data directly into the data stream connected to the database. Such a threat is shown as AP 7 and AP 10 in Fig. \ref{fig:ourmodel}. The data transmitted over the communication channel can be encrypted to avoid interception.

\subsubsection{Modify score}
The adversary can ease the process of authentication by modifying the matching score intercepted between the channel connecting the matcher and the decision module. He overrides the actual score with a high score above the threshold. Fig. \ref{fig:ourmodel} represents such a threat as AP 12. A mutual authentication between matcher and decision module, like a challenge-response technique, can be used to detect a modified score.

\subsection{Modify access rights}
The administrator enrolls users having different privileges or access rights in the biometric controlled application. If the adversary diminishes the rights of a high privileged user, then it leads to a denial of service attack to such users. The system is under threat when a user registered with low access rights suddenly gets administrative rights. In either case, the adversary initially needs to gain the system administrator credentials. 

\subsection{Decision override}
The decision module accepts or rejects a user as enrolled and authorized based on the comparison between the matching score and the predefined score. The attacker can use a Trojan horse to override the decision module to give the result in his favor. For example, such an attack can be used by the adversary to grant access to the application in Fig. \ref{fig:ourmodel} or deny all other users from using the application. 

\subsection{Denial-of-service (DoS) attack}
The DoS attack renders the system inaccessible to the authorized users. For a biometric authentication system, an online authentication server that processes access requests (via retrieving templates from a database and performing matching with the transferred biometric data) can be bombarded with many bogus access requests to a point, where the server's computational resources cannot handle valid requests any more \cite{DBLP:conf/sswmc/UludagJ04}. These attacks are generally detected in  a small time duration. In some cases, however, the intent is to have the attack noticed in order to create confusion and alarm and force the activation of alternative or exception handling procedures \cite{Roberts:2007:BAV:2639537.2639843}. The organization can use CCTV cameras, tampering alarms, or personal security guards to protect a biometric system controlled application, such as a cash dispenser shown in Fig.\ref{ourmodel}.

\subsection{Intrusion}
In intrusion attack, the attacker breaks into the template storage through an external system. The scenario where the system recognizes an impostor as an authorized user is also a case of intrusion. The server hardening (i.e. enhancing the server security against all possible vulnerabilities), database access controls, signing  stored templates, storing encrypted templates and use of Match-on-Card technology can be the possible countermeasures against intrusion attack.

\subsection{Zero-effort attacks}
The zero-effort attack is a benefit an adversary receives, and the trouble an enrolled user faces due to system limitations. It is an incorrect authentication (a non-enrolled user is authenticated due to false match error) and rejection (an enrolled user is rejected due to false non-match error) scenario. This attack is the result of shifting the predefined threshold to lower or higher value. Fig. \ref{fig:ourmodel} shows this scenario as AP 14 and AP 15. The existence of zero-effort attack is due to the limited individuality and intraclass variations in biometric features and efforts are underway by the research community to reduce such intrinsic errors by designing salient-feature detectors and robust matchers \cite{DBLP:journals/ejasp/JainNN08}.

An on-line platform, EvaBio, for the security evaluation of biometric systems was presented in \cite{DBLP:conf/icb/El-AbedLR12}, which implements a quantitative-based security assessment method to allow easily the evaluation and comparison of biometric systems. In order to quantify a newly developed biometric system, the authors also provided a database of common vulnerabilities and threats of a biometric system. 

\begin{table*}[t]
	\centering
	\caption{Biometric system attack summary with reference to Fig. 4. AP stands for Attack Point.}
	\label{attackTable}
	\begin{tabular}{|c|c|c|c|c|}
		\hline
		\textbf{AP} & \textbf{Targeted component}                                                                                                                  & \textbf{Possible attack} 
		& \textbf{Countermeasures}
		& \textbf{References} \\ \hline
		1                     & Biometric input device                                                                                                                       & \begin{tabular}[c]{@{}c@{}}Adversary attacks and \\ administrative frauds\end{tabular}    & \begin{tabular}[c]{@{}c@{}}Liveness detection, \\ Challenge/response \end{tabular} & \begin{tabular}[c]{@{}c@{}}\cite{Adler2008}, \cite{DBLP:journals/prl/GalballyCLRMFOM10}, \cite{DBLP:journals/spm/HadidEMF15}, \\ \cite{DBLP:journals/csur/MarascoR14}, \cite{DBLP:conf/icpr/FengJR10} \end{tabular}  \\ \hline
		2                     & Biometric input device                                                                                                                       & Denial of service                                                                         & Rugged devices  & \cite{Roberts:2007:BAV:2639537.2639843}, \cite{Jain:2008:BTS:1376536.1387883}    \\ \hline
		3                     & \begin{tabular}[c]{@{}c@{}}Communication channel between sensor \\ and feature extraction module\end{tabular}                                & Fingerprint image intercept                                                               & \begin{tabular}[c]{@{}c@{}}Transmit data over \\encrypted path/secure channel\end{tabular} & \cite{Jain:2008:BTS:1376536.1387883}    \\ \hline
		4                     & Feature extraction module                                                                                                                    & \begin{tabular}[c]{@{}c@{}}Trojan horse attack,\\ Override feature extractor\end{tabular} & Code signing & \cite{Jain:2008:BTS:1376536.1387883}    \\ \hline
		5                     & \begin{tabular}[c]{@{}c@{}}Communication channel between feature extraction \\ module and template protection techniques module\end{tabular} & Replay attack                                                                             & \begin{tabular}[c]{@{}c@{}}Challenge-response \\based system, \\disposable Feature Extractors (FE)\end{tabular}& \cite{Roberts:2007:BAV:2639537.2639843}, \cite{DBLP:journals/procedia/SheltonBASALARD12}    \\ \hline
		6                     & Template protection techniques module                                                                                                        & Side channel attacks                                                                      & \begin{tabular}[c]{@{}c@{}}Masking, designing ICs \\with active shield \end{tabular}& \cite{DBLP:conf/ccs/DurmuthOP16}, \cite{Tiri2005ASL}   \\ \hline
		7                     & \begin{tabular}[c]{@{}c@{}}Communication channel between template protection \\ techniques module and template database\end{tabular}         & \begin{tabular}[c]{@{}c@{}}Template intercept,\\ data inject\end{tabular}                 & \begin{tabular}[c]{@{}c@{}}Use strong tested \\biometric algorithms \end{tabular}& \cite{604477}    \\ \hline
		8                     & Template database                                                                                                                            & \begin{tabular}[c]{@{}c@{}}Steal, delete, modify,\\ substitute template\end{tabular}      & \begin{tabular}[c]{@{}c@{}}Hardened server, \\DB access controls, \\ sign templates, \\store encrypted templates,\\ store template  on card\end{tabular}& \cite{DBLP:journals/ejasp/JainNN08}, \cite{DBLP:journals/ieeesp/PrabhakarPJ03}, \cite{DBLP:conf/eusipco/JainRU05}   \\ \hline
		9                     & \begin{tabular}[c]{@{}c@{}}Communication channel between template protection \\techniques module and matcher module\end{tabular}            & \begin{tabular}[c]{@{}c@{}}Hill-climbing attack,\\ brute force attack\end{tabular}        & Time out/lock out policies & \cite{4105331}, \cite{DBLP:journals/corr/abs-1304-7386}   \\ \hline
		10                    & \begin{tabular}[c]{@{}c@{}}Communication channel between template database\\  and matcher module\end{tabular}                                & Replay attack                                                                             & \begin{tabular}[c]{@{}c@{}}Utilize Timestamps or \\Time to Live (TTL) tag \end{tabular} & \cite{Roberts:2007:BAV:2639537.2639843}    \\ \hline
		11                    & Matcher module                                                                                                                               & \begin{tabular}[c]{@{}c@{}}Side channel attack,\\ Trojan horse attack\end{tabular}        & \begin{tabular}[c]{@{}c@{}}Masking, designing ICs \\with active shield, \\Code signing \end{tabular}  & \cite{DBLP:conf/ccs/DurmuthOP16}, \cite{Yang:2003:SFM:982507.982524}   \\ \hline
		12                    & \begin{tabular}[c]{@{}c@{}}Communication channel between matcher module and \\ decision module\end{tabular}                                  & Modify score                                                                              & \begin{tabular}[c]{@{}c@{}}Mutual authentication between \\matcher and decision module \end{tabular} & \cite{Jain:2008:BTS:1376536.1387883}    \\ \hline
		13                    & Decision module                                                                                                                              & Override decision                                                                         & Code signing & \cite{Roberts:2007:BAV:2639537.2639843}    \\ \hline
		14                    & \begin{tabular}[c]{@{}c@{}}Communication channel between decision module and \\ biometric application\end{tabular}                           & \begin{tabular}[c]{@{}c@{}}Zero-effort attack\\ (false match error)\end{tabular}          & Design robust matcher & \cite{DBLP:journals/corr/abs-1304-7386}, \cite{Jain:2006}    \\ \hline
		15                    & \begin{tabular}[c]{@{}c@{}}Communication channel between decision module and \\ biometric application\end{tabular}                           & \begin{tabular}[c]{@{}c@{}}Zero-effort attack\\ (false nonmatch error)\end{tabular}       & Design robust matcher  & \cite{Jain:2006}    \\ \hline
		16                    & Biometric application (e.g. cash dispenser)                                                                                                  & Denial of service                                                                         & \begin{tabular}[c]{@{}c@{}}CCTV monitoring, \\deploy security guards  \end{tabular} & \cite{Jain:2008:BTS:1376536.1387883}, \cite{Roberts:2007:BAV:2639537.2639843}    \\ \hline
	\end{tabular}
\end{table*}

\section{Threat models}\label{models}

In this Section, we present threat models that exist for a fingerprint biometric system. A threat model is a way to identify various possibilities of attacks (or the potential threats) on a system in the computational powers of an adversary. We present four threat models that exist in the literature which point out different attacks and vulnerabilities in the biometric system.

\subsection{Ratha et al. model}

\begin{figure}[b]
	\centering	
	\includegraphics[width=0.30\textwidth]{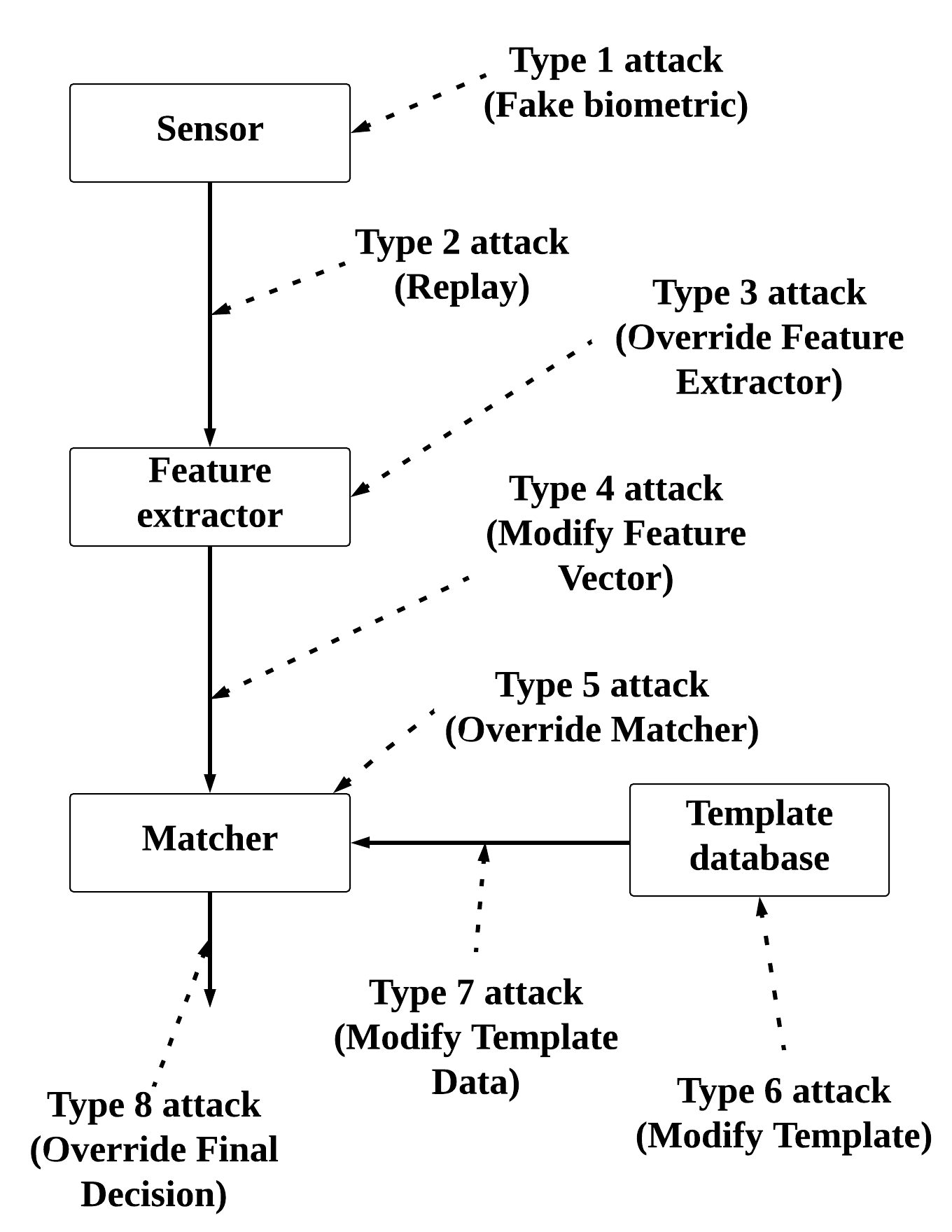}
	\caption{Ratha et al. model \cite{Ratha} : The dashed arrows indicate the access point of information for mounting a specific type of attack}
	\label{fig:ratha}
\end{figure}

The \textit{Ratha et al. model} shown in Fig. \ref{fig:ratha} is the first threat model proposed for a fingerprint biometric system. It identifies eight vulnerabilities based on the access point of information in a biometric system \cite{Ratha}. These vulnerabilities lead to a specific type of attack, represented as Type 1 to Type 8, on the system components. The countermeasures for all 8 types of attacks from this model are discussed in Section \ref{attacks}.

\textit{Type 1 attack} : According to the model, sensors are susceptible to spoofing, false biometric submission or residual attack represented as Type 1 attacks. These attacks are basically a fake biometric presentation at the sensor. The adversary collects the biometrics of a genuine user and creates fake physical finger to access the system. 

\textit{Type 2 attack} : The interception of communication channel between sensor and feature extractor by the adversary forms the base for Type 2 attack. The adversary replays a previously intercepted digitized biometric signal bypassing the sensor. Masquerading is a Type 2 attack.

\textit{Type 3 attack} : The attacker can override the feature extraction module to mount a Type 3 attack. The adversary copies a Trojan horse executable at the feature extraction module. Generally, Trojan horses can be controlled remotely. The adversary sends commands to the Trojan horse to send the feature set selected by him.

\textit{Type 4 attack} : The communication channel between the feature extractor and the matcher can be targeted by the attacker. The adversary intercepts the features transmitted over the communication channel and replays it after a time gap. The previously intercepted and tampered/modified feature sets can be directly transmitted to the matcher bypassing the feature extractor module. 

\textit{Type 5 attack} : The threat to the matcher module overriding the matching score using a Trojan horse is a Type 5 attack. In this attack, the adversary sends commands remotely to generate a high matching score and send positive response (e.g. a ``Yes'' or ``Accept'') to the biometric controlled application  and bypass the authentication process completely. He can even mount a denial-of-service attack by giving a command to indefinitely generate a low matching score. 

\textit{Type 6 attack} : The intruder can explore ways for template database leakage with Type 6 attack. The adversary not only collects the leaked stored templates but also replays them and modifies them to get authorized on behalf of different enrolled users. 

\textit{Type 7 attack} : In Type 7 attack scenario, the adversary intercepts the communication channel between the template database and the matcher module to collect templates from the data stream. These intercepted templates can be directly replayed or altered and then replayed at the matcher module to access the system with the identity of different users.

\textit{Type 8 attack} : The adversary can bypass all system components and directly manipulate the system decision in his favor in Type 8 attack. An attack to override the final decision of the biometric system in favor of the adversary nullifies the excellent performance characteristics of the biometric application.

\begin{figure}[b]
	\centering	
	\includegraphics[width=0.5\textwidth]{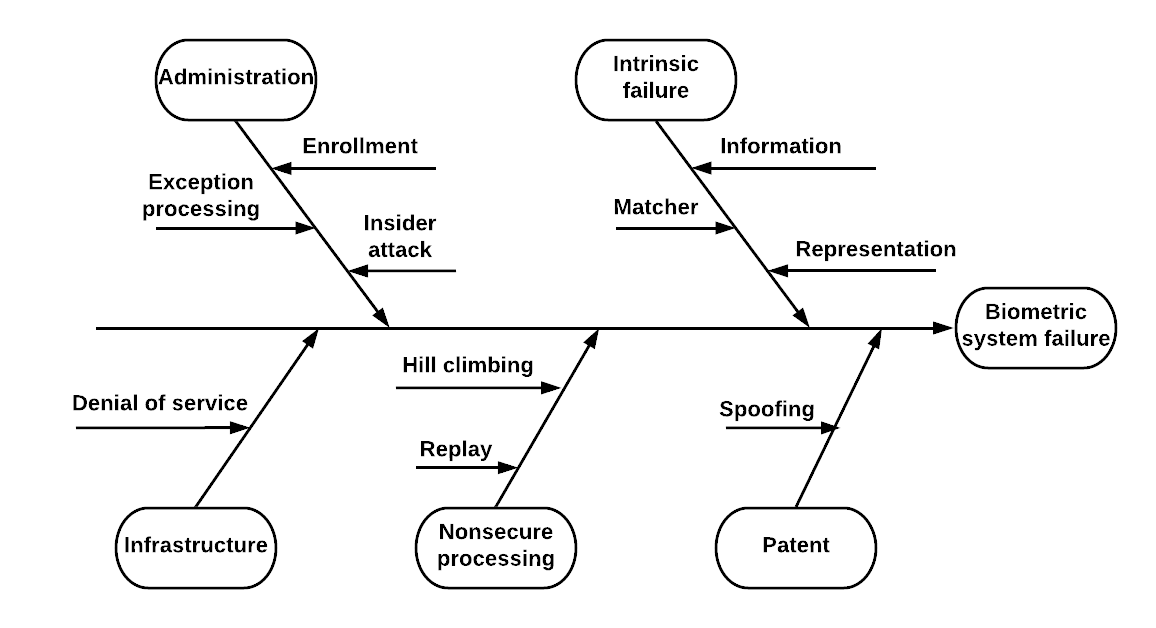}
	\caption{Fishbone model \cite{Jain:2008:BTS:1376536.1387883}. The blocks denote the processing phases of a biometric system that are vulnerable targets in this model. The possible attacks are mentioned on the arrows. 
	}	
	\label{fig:fishbone}
\end{figure}

\begin{figure}[!ht]
	\centering	
	\includegraphics[width=0.5\textwidth]{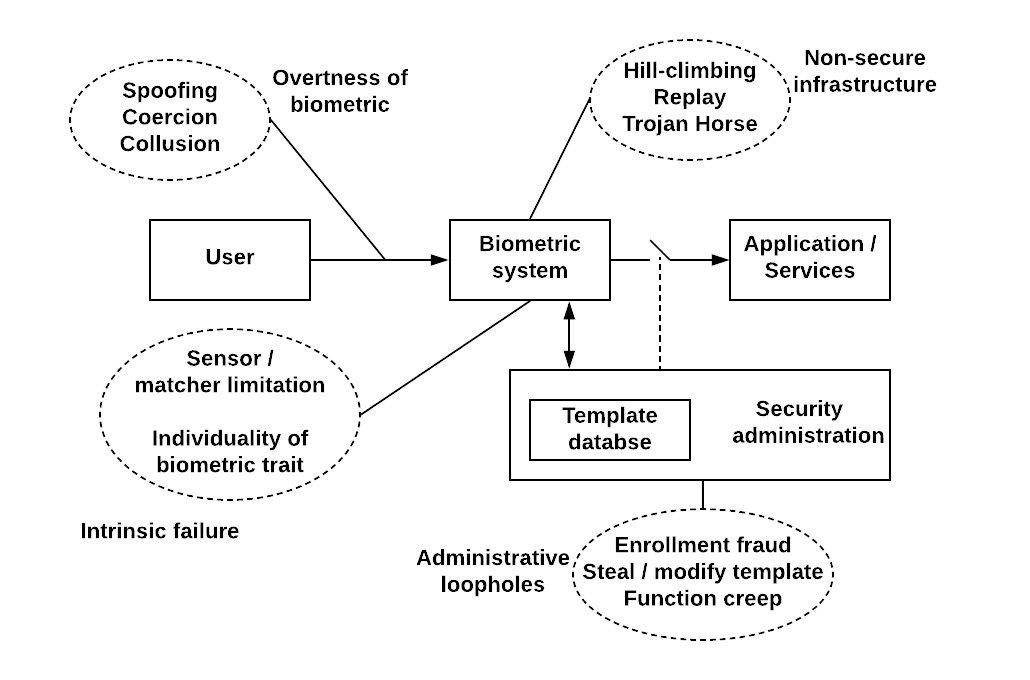}
	\caption{Nagar et al. model \cite{Jain:2008:BTS:1376536.1387883}}	
	\label{fig:nagar}
\end{figure}

\begin{figure*}
	\includegraphics[width=\textwidth,height=17cm]{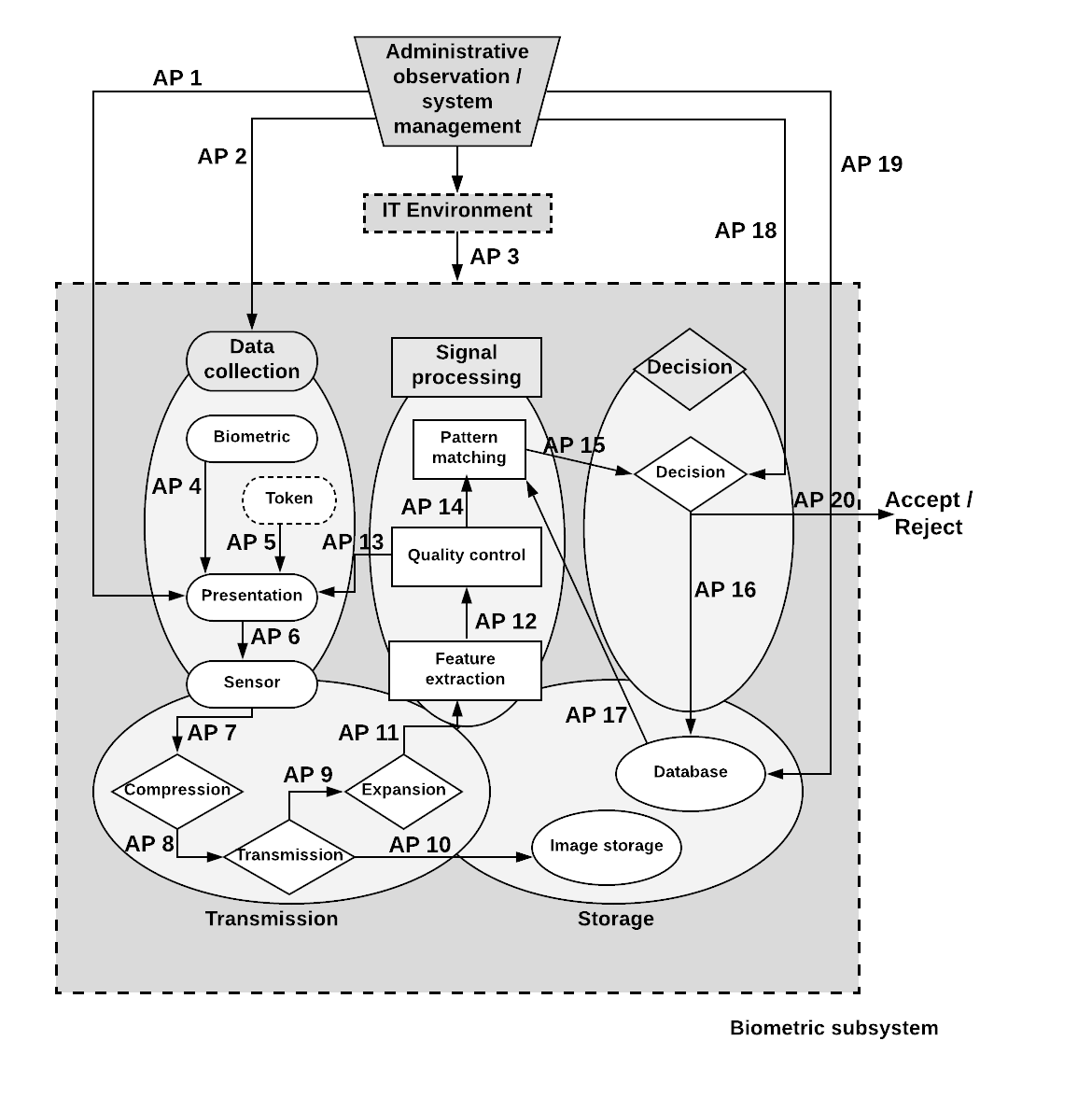}
	\caption{Bartlow and Cukic framework : AP represents an Attack Point. The biometric system is divided into three modules, namely, administrative observation/system management module, IT environment module and biometric subsystem module. The biometric subsystem module is divided into five subsystem,  namely, data collection, signal processing, decision, transmission and  storage subsystem.}
	\label{fig:bartlow}
\end{figure*}

\subsection{The fishbone model}
Fig. \ref{fig:fishbone} shows the \textit{fishbone} (cause and effect) model proposed by Jain et al. in \cite{Jain:2008:BTS:1376536.1387883}. The model depicts the causes that determine the vulnerability of a biometric system. The authors identified five causes that lead to vulnerabilities in the biometric system represented in ovals in Fig. \ref{fig:fishbone}. The arrows specify the effects i.e. the possible attacks due to these causes.

\textit{Intrinsic failure} involves the system limitations and errors responsible for false acceptance of a non-enrolled individual or false rejection of an enrolled user. The intra-user variations causes large variations in the live and stored features of the enrolled user leading to false rejection. This error can be due to improper or partial finger put at the sensor. The lack of individuality or uniqueness (mostly among twins in face recognition system) can result in large similarity between the feature sets of two different individual leading to false rejection. 

\textit{Administration cause} describe the ways in which a dishonest staff, such as, a system administrator can play the role of an adversary. A disloyal administrator can assist an adversary in either mounting a successful attack, or induce an enrollment fraud or exception abuse. The exception processing is a way for the enrolled user to get authorized in case if the biometric system rejects him due to intrinsic limitation or denial of service attack (by deleting his stored template). The insider attacks include the cooperation an adversary receives from the disloyal staff to mount an attack on the system. 

\textit{Infrastructure causes} include system design defects that make the system susceptible to adversary attacks. The hardware components such as sensor, the software implemented at the feature extraction module and matcher module along with the communication channel between various system components forms the infrastructure of the biometric system. A Trojan horse program can be used by the adversary to override the matcher or decision module. 

\textit{Non-secure processing causes} point out the vulnerabilities due to insecure enrollment and authentication process. The adversary can mount a hill-climbing or replay attack through the non-secure communication channel to intercept the information and get authorized by skipping the sensor. The poorly secured database can leak the stored templates. The adversary uses the templates thus collected to mount a replay attack. A function creep is a phenomenon of cross-matching the biometric features of an enrolled user. The adversary collects the biometrics of a user either through database leakage or communication channel intercept from his employer's system and uses it in other applications like banking services, passport services etc.

\textit{Patent cause or biometric overtness} is related to the secrecy of biometric templates. The attacker covertly collects the biometrics of a genuine user from the sensor or public places visited by the user. These biometrics can be used to create fake physical or gummy finger to mount a spoofing attack. 

This model highlights the general errors that should be avoided and the security techniques to be implemented while designing a biometric system. The authors suggest that the most straightforward way for securing the biometric system is to store the template and the system modules (components) on smart cards. Such systems are called as system-on-card or match-on-card.

\subsection{The Nagar et al. model}
Fig. \ref{fig:nagar} shows Nagar et al. model. This model is based on the Fishbone model in terms of specifying causes for vulnerabilities and their effects on a biometric system. The text besides the dotted ovals in Fig. \ref{fig:nagar} specify a given cause and the contents of the ovals shows the possible attacks as an effect of these causes. The four large rectangles in Fig. \ref{fig:nagar} shows the major constituents of a biometric-based authentication application. 

The \textit{non-secure infrastructure} can leak sensitive information to the adversary through communication channel. The system components, such as feature extraction module or the matcher module are also vulnerable to Trojan attack which overrides the real output of these elements. The authors combined non-secure processing causes (arising due to insecure enrollment and authentication process) and infrastructure causes (responsible for denial of service attack) from Fishbone model into a single non-secure infrastructure cause in this model. 

\textit{Overtness of biometric} includes the attacks that use the overt information about the biometrics of an individual, such as, our face, fingerprint left at public places, our voice over a conversation etc. The attack involving overtness of biometric include coercion, collusion, and spoofing attacks targeted by the adversary at a biometric input device.

\textit{Intrinsic failure} includes the system limitations responsible for false acceptance of a non-registered user. The predefined threshold decides the authenticity of a claiming user. For a high security application, like entry into military zone, the threshold is fixed to a higher value which sometime may reject a soldier due to improper touch of his finger to the sensor or environmental factors. For a university attendance system, a student not registered for a particular course can mark proxy attendance for a registered student due to lower threshold value.

A system administrator may exploit \textit{administrative loopholes} to misuse his administrative rights. As discussed in Section \ref{attacks}, the administrator can support the adversary in collusion, enrollment fraud or spoofing. The best countermeasure against the attacks involving internal staff with administrative privileges is to appoint multiple system administrators for every superuser task.

\subsection{Bartlow and Cukic framework}
Wayman proposed to classify a biometric system as a composition of five primary subsystems depending on its application, namely, data collection, transmission, signal processing (which includes feature extraction, quality control, pattern matching), storage, and decision \cite{Wayman1996}. Fig. \ref{fig:bartlow} shows the Bartlow and Cukic framework \cite{Bartlow}, \cite{Roberts:2007:BAV:2639537.2639843} as an extension of Ratha et al. model and Wayman's subsystem architecture, which mainly focus on technical testing of biometric devices. The authors decomposed the model into three modules and identified more than twenty vulnerable attack points. The three modules (shown in dark gray color in Fig. \ref{fig:bartlow}) are namely administrative observation/system management module, IT environment module, and the biometric subsystem module. The biometric subsystem (i.e. the system management module) is further categorized into five subsystems (shown in large ovals in Fig. \ref{fig:bartlow}) according to Wayman's suggestion. The arrows represent the point of attacks and vulnerabilities in the subsystems and various modules in the system.

In Fig. \ref{fig:bartlow}, the \textit{administrative observation/system management} module represents disloyal system administrators and the arrow originating from it shows the possible attack points by such staff targeting different components. Attack points AP 1 and AP 2 target the data collection module of biometric subsystem where the adversary seeks help from the administrator. Attack points AP 18 and AP 20 specify the manipulation of final decision by the administrator in favor of the adversary. The bad administrator can perform false enrollment through AP 19.

\begin{table*}[h]
	\centering
	\caption{Comparison of Threat Models}
	\label{threat_table}
	\begin{tabular}{|c|c|c|c|}
		\hline
		\textbf{Threat model}                                                  & \textbf{Underlying criteria}                                                         & \textbf{Classification}                                                                           & \textbf{Application}                                                                                                                     \\ \hline
		Ratha et al. model                                                     & \begin{tabular}[c]{@{}c@{}}Access point of \\ information\end{tabular}               & \begin{tabular}[c]{@{}c@{}}8 vulnerable points resulting \\in 8 types of attacks\end{tabular} & \begin{tabular}[c]{@{}c@{}}A general model without \\details of specific attacks\end{tabular}                                      \\ \hline
		Fishbone model                                                         & cause and effect                                                                     & \begin{tabular}[c]{@{}c@{}}5 causes leading to\\ biometric system failure\end{tabular}            & \begin{tabular}[c]{@{}c@{}}Useful while designing\\ biometric security techniques\end{tabular}                                    \\ \hline
		Nagar et al. model                                                     & \begin{tabular}[c]{@{}c@{}}Cause and effect similar\\ to Fishbone model\end{tabular} & \begin{tabular}[c]{@{}c@{}}Major vulnerabilities and \\ their 4 underlying causes\end{tabular}    & \begin{tabular}[c]{@{}c@{}}Specifies the vulnerabilities for\\ biometric-based authentication system\end{tabular} \\ \hline
		\begin{tabular}[c]{@{}c@{}}Bartlow and Cukic \\ framework\end{tabular} & \begin{tabular}[c]{@{}c@{}}Wayman's subsystem\\ architecture \cite{Wayman1996} \end{tabular}            & \begin{tabular}[c]{@{}c@{}}3 modules with\\ 20 potential attack points\end{tabular}               & \begin{tabular}[c]{@{}c@{}}Well suited for testing\\ and validating secure techniques\end{tabular}                                \\ \hline
	\end{tabular}
\end{table*}

The \textit{IT environment subsystem} includes other applications, such as operating system and database management system that directly or indirectly interact with the biometric system. The attack point AP 3 shows the possibility of an attack through these interacting applications. There is a possibility of intrusion or malicious code execution with the help of these applications.

The \textit{biometric subsystem} specifies the internal components of the biometric system, the data transmission between them, and the targeted vulnerable points. In the \textit{data collection} module, the collection and presentation of biometrics is performed. The attack points, AP 4, AP 5, and AP 6 are vulnerable to attacks carried out through the input devices. The spoofing, replay, and masquerade attacks can be mounted on this module. The compression and transmission of biometric data is represented with \textit{transmission module} which is susceptible to attacks, such as replay, parallel sessions, and masquerade shown by attack points, AP 7, AP 8, AP 9, AP 10, and AP 11 in Fig. \ref{fig:bartlow}. In this attack mode, the adversary intercepting over the communication channel collects the information required to mount replay, hill-climbing, or brute force attack. The \textit{signal processing} module comprises the robust feature extraction, template matching, and image quality control components. The adversary uses a Trojan at AP 12 and AP 14 to target the feature extractor and the quality control unit, respectively. The attacker can even use a poor quality image at AP 13 to mount a hill-climbing attack, or execute a brute force attack. Using the hill-climbing attack, the attacker can construct a fingerprint image using the responses (i.e. score) of matcher module until an acceptable score is achieved. In another scenario, he can apply image processing techniques to generate multiple copies of the poor quality image by enhancing the ridges and use all these images to mount a brute force attack. The interception of information between the pattern matching module and the decision module is shown in AP 15. The process of capturing data through database leakage and retransmitting it again to the matcher module is done at attack points, AP 16 and AP 17, respectively. The attempt to override the final conclusion of the \textit{decision} module in cooperation with a bad administrator is represented with AP 20.

The details about potential vulnerable attack points is given in this model, but it does not propose the methodology to design security techniques for a biometric system. So, this model can be used only as a benchmark to test and validate existing and proposed security techniques.

\section{Template protection schemes}\label{template}
A template represents a set of salient features, such as minutia, that summarizes the biometric data (signal) of an individual \cite{DBLP:conf/eusipco/JainRU05}. The stored template should not reveal any data that can be replayed and it should be difficult for an adversary to guess or reverse engineer the original biometric trait or any close replica from the stored data \cite{DBLP:journals/ejasp/JainNN08}. The biometric template protection schemes aim at preserving the user's privacy while enhancing the security of stored templates.

\begin{figure*}[t]
	\includegraphics[width=\textwidth,height=13cm]{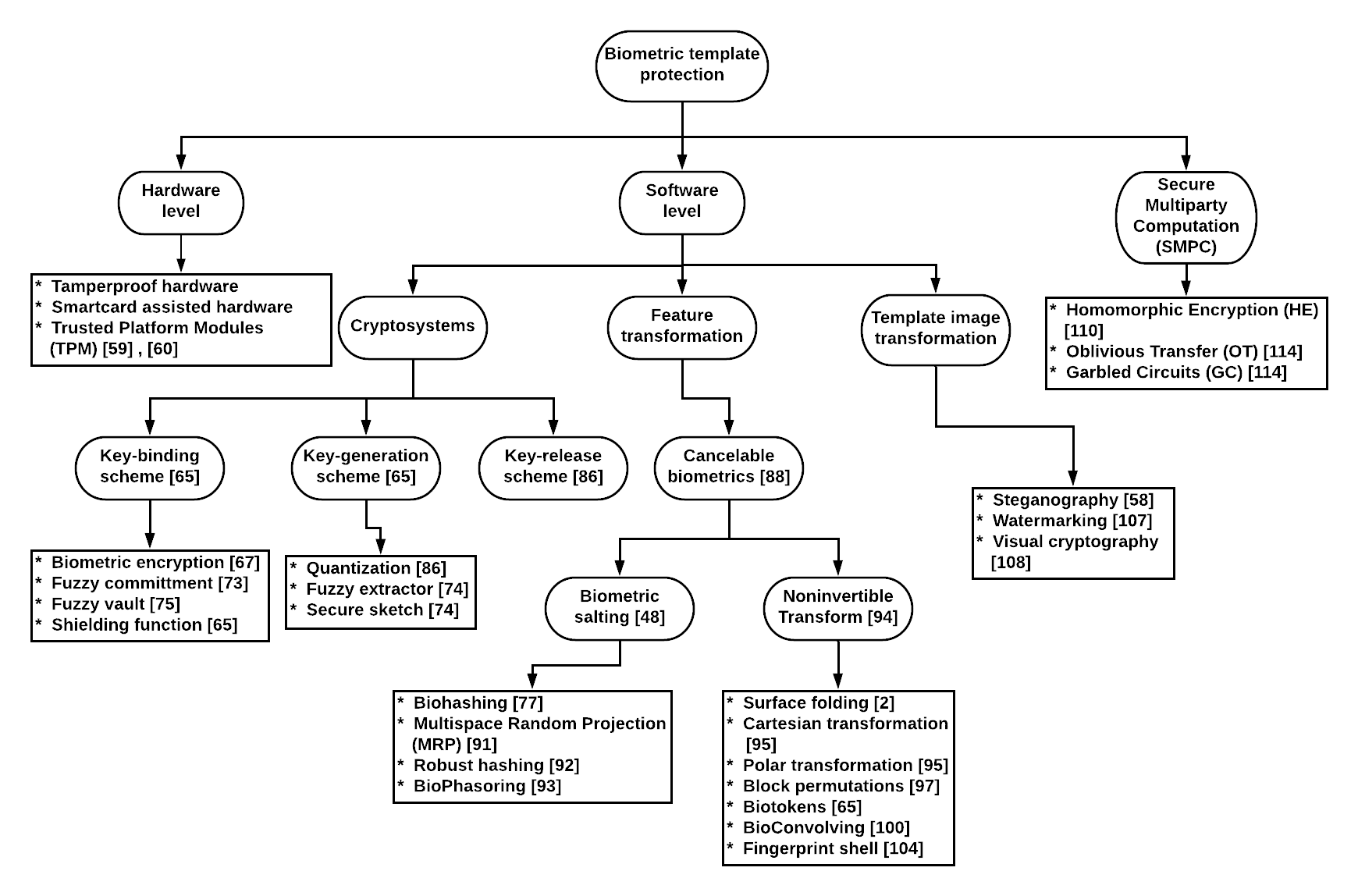}
	\caption{Template protection schemes}
	\label{fig:template}
\end{figure*}

ISO/IEC 24745 \cite{ISO24745} standard specifies three mandatory requirements for all biometric template protection schemes, namely, non-invertability, revocability, and unlinkability. \textit{Non-invertability} property requires that given a protected biometric template, it should be computationally infeasible to obtain the original template. \textit{Revocability} property requires that, it should be computationally difficult to obtain the original biometric template from multiple instances of protected biometric reference derived from the same biometric trait of an individual \cite{DBLP:journals/spm/NandakumarJ15}. \textit{Non-linkability} property assures that, it is computationally difficult to ascertain whether two or more instances of protected biometric reference were derived from the same biometric traits of a user \cite{DBLP:journals/spm/NandakumarJ15}.

Fig. \ref{fig:template} shows various template protection schemes adapted from \cite{DBLP:journals/mta/KaurK16}. Biometric templates storing minutia information (without encryption) are vulnerable to tampering attack and replay attack. The templates can be secured at the database (by avoiding leakage), over the communication channel (by using encryption techniques), and also at the matcher (by concealing any information related to secret data, such as encryption key, through data-dependent variations in power consumption or computation time). The schemes to protect biometric templates can be broadly categorized into three levels, namely, hardware level, software level, and biometrics in encrypted domain.

\subsection{Hardware level schemes}
Hardware level protection to templates is performed during transmission and in the database to avoid interception and tampering of templates. A \textit{tamper-proof hardware} disallows an  attacker from breaking into the hardware components, such as matcher, to steal a secret (such as, an encryption key), and thus prevent unauthorized access. \textit{Smart-card assisted hardware}, such as System-on-Card and Match-on-Card, can help in mitigating such attacks on biometric templates. Since these systems embeds the biometric system and a user template on the card itself, it lowers the probability of template interception and leakage from database as compared to online database. The \textit{trusted platform module} (TPM)  was developed by Trusted Computing Group (TCG) as platform  that includes additional hardware and software to increase the security level of IT-systems \cite{TPM1}. A Trusted Platform is ``a computing platform that has a trusted component, probably in the form of built-in hardware, which it uses to create a foundation of trust for software processes'' \cite{pearsontpm}. TPM secures a hardware (i.e., a cryptoprocessor) using an integrated cryptographic key. TPM can be used for hardware level tamper detection and also for hardware-based authentication. There are increased use of TPM chips in certain dedicated security applications, such as Microsoft BitLocker, storage of public key infrastructure (PKI), private keys, and other credentials (e.g., biometric identifiers), and potentially secure machine identities on sensitive networks \cite{tpmexample}.

\iffalse 
A \textit{leakage proof hardware} guarantees that an adversary is prohibited from attempting a side-channel attack. Such hardware does not leak information through power consumption, computation time etc. Tiri et al. have presented a secure co-processor (built in a $0.18\mu m$ CMOS technology) that does not leak information through the power supply, which is a major and easy to access side-channel leakage source and believed to be the first IC that is practically immune to DPA attacks \cite{Tiri2005ASL} 
\fi

\subsection{Software level schemes}
The software level schemes mainly concentrate on the ways to make sure that the biometrics stored on the templates is encrypted (using key) and it is practically infeasible to divulge the secret encryption key or regenerate the original fingerprints of a user. These schemes include cryptosystems (using encryption key to secure template), template generation using feature transformations (such as, salting and non-invertible transforms), and transformation of template image (e.g. steganography).

\subsubsection{Cryptosystems}
Biometric cryptosystems provide the benefits of cryptography and biometrics to an application. They protect the application using high and adjustable security levels provided by cryptographic techniques using a secret key, and ensures trust using non-repudiation through biometrics. The biometric cryptosystems can be designed in two ways. It can either use the biometrics of an individual to generate a digital key, or securely bind a secret key to the user biometrics. Biometrics-based ``key-release'' refers to the use of biometric authentication to release a previously stored cryptographic key, whereas biometrics-based ``key generation'' refers to extracting/generating a cryptographic key from a biometric template or construct \cite{DBLP:conf/iscis/KholmatovY06}. These techniques are discussed in detail in the following paragraphs of this Section.

The majority of cryptosystems require the storage of biometric dependent public information, which is applied to retrieve or generate keys; such biometric dependent information is referred to as {\em helper data} \cite{Jain:2008:BTS:1376536.1387883}. Uludag et al. \cite{DBLP:conf/cvpr/UludagJ06} stated that two contradicting requirements arise,  
\begin{enumerate}
	\item helper data should contain sufficient information to allow an alignment between a query and a template, and 
	\item helper data should not leak critical information about the template. 
\end{enumerate}
The authors proposed a scheme for generating orientation-field based helper data  in the {\em fuzzy fingerprint vault} framework, which does not leak minutiae information, yet it enables us to properly align the input query.

\subsubsection*{Key-binding scheme}
In biometric-based key binding schemes, the cryptographic key is tightly bound with the biometric template so that it
cannot be released without a successful biometric authentication without accessing template directly \cite{DBLP:conf/IEEEias/LiWPZ09}. The transformed characteristic of the legitimate user is compact in the feature space, and the characteristic of a fake identity is either dispersed or far away from the legitimate user's characteristic. This variation distinguishes the transformed characteristic of the registered user from the characteristic of the false identity.

\begin{figure}[h]
	\centering
	\includegraphics[width=0.50\textwidth]{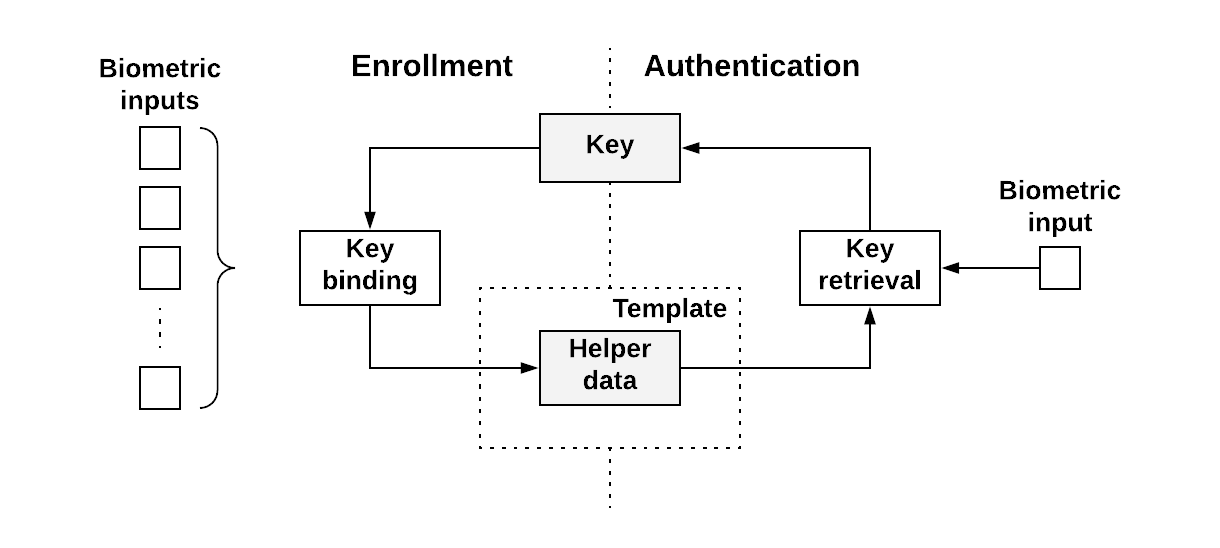}
	\caption{Basic concept of biometric key-binding \cite{DBLP:journals/ejisec/RathgebU11}.}
	\label{fig:keybinding}
\end{figure}

Fig. \ref{fig:keybinding} shows the basic concept behind the key-binding scheme. In this method, the key is combined with a biometric sample to generate helper data during the enrollment. A new biometric sample along with the stored helper data are used to release the same key during the authentication \cite{DBLP:journals/eswa/ImamverdiyevTK13}. In this scheme, it is computationally infeasible to retrieve the original biometric template or the secret key given only the helper data.

\textit{Biometric Encryption} (BE) is a group of emerging technologies that securely binds a digital key to a biometric or generates a digital key from the biometric so that no biometric image or template needs to be stored \cite{DBLP:reference/bio/CavoukianS15}. In biometric encryption, a password-based encryption key is used to encrypt the template during the enrollment phase. The most distinct BE technologies are the following: Mytec1 \cite{SouterPatent} and Mytec2 \cite{doi:10.1117/12.304770}, \cite{doi:10.1117/12.304705}, ECC check bits \cite{DBLP:conf/wifs/WangRV09}, biometrically hardened passwords \cite{DBLP:journals/ijisec/MonroseRW02}, the fuzzy commitment scheme \cite{DBLP:journals/tifs/IgnatenkoW10} and some of its generalizations in the fuzzy extractor/secure sketch framework \cite{DBLP:journals/corr/abs-cs-0602007}, shielding functions (i.e., quantization using correction vector) \cite{DBLP:conf/IEEEias/LiWPZ09}, fuzzy vault \cite{DBLP:journals/iacr/JuelsS02}, PinSketch \cite{DBLP:conf/eurocrypt/DodisRS04}, and BioHashing \cite{DBLP:conf/biosig/TopcuEKY13} with key binding \cite{doi:10.1002/9780470522356.ch26}.

Biometric Encryption\textsuperscript{TM}, an algorithm for linking and retrieval of digital keys, which can be used as a method for the secure management of cryptographic keys was proposed by Soutar et al. in \cite[Chapter~22]{BiometricEncryption}. This algorithm works on the complete fingerprint image and is based on the mechanism of correlation. The algorithm independently generates the cryptographic key which can be periodically updated using a re-enrollment procedure. 

During enrollment, the Biometric Encryption\textsuperscript{TM} process combines the biometric image with a digital key (used as a cryptographic key) to create a secure block of data, known as a Bioscrypt\textsuperscript{TM} \cite[Chapter~22]{BiometricEncryption}. Mytec Technologies Inc. claim that it is infeasible to obtain either the fingerprint or the key independently from Bioscrypt \textsuperscript{TM}. The algorithm combines the biometric image and the Bioscrypt\textsuperscript{TM} to retrieve the key during verification. 

In the enrollment process, a set of input fingerprint images undergo image processing with a random (phase) array and generate two new arrays as output, namely, $H_{stored}(u)$ which is stored in Bioscrypt\textsuperscript{TM} and $c_{0}(x)$. The proprietary link algorithm performs linking of cryptographic key $k_{0}$ and $c_{0}(x)$. Finally, the key $k_{0}$ is used to create an identification code $id_{0}$. During verification process, the $H_{stored}(u)$ undergoes image processing with a new set of input (claim) fingerprint images to produce an output pattern $c_{1}(x)$. The proprietary retrieval algorithm accepts $c_{1}(x)$ as input to extract a key $k_{1}$. A new identification code $id_{1}$ is generated from $k_{1}$. In the end, the key validation is performed by comparing $id_{0}$ and $id_{1}$.

\begin{figure}[h]
	\centering
	\includegraphics[width=0.50\textwidth]{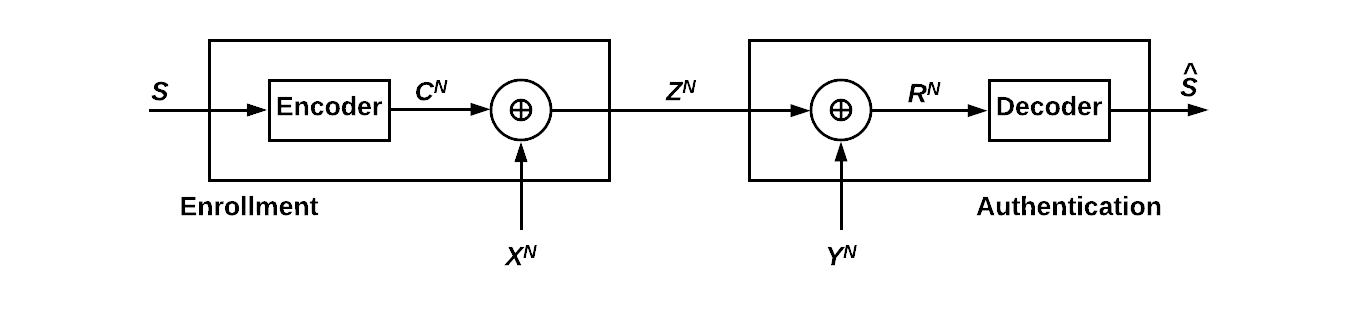}
	\caption{Fuzzy commitment scheme \cite{DBLP:journals/tifs/IgnatenkoW10}}
	\label{fig:fuzzycommitment}
\end{figure}

\textit{Fuzzy commitment}
A fuzzy commitment scheme (FCS) processes a binary biometric enrollment sequence (i.e. a binarized fingerprint image) $X^{N}=\{x_{1},x_{2},...,x_{N}\}$ with symbols, $x_{i}\in\{0,1\}$ for $i=\{1,2,...,N\}$, and a binary biometric authentication sequence $Y^{N}=\{y_{1},y_{2},\ldots,y_{N}\}$ with symbols $y_{i}\in\{0,1\}$ for $i=\{1,2,...,N\}$, where the sequences are generated by a biometric source according to some distribution $\{Q( \,x^{N},y^{N}) \,,x^{N} \in \{0,1\}^{N},y^{N} \in \{0,1\}^{N} \}$ \cite{DBLP:journals/tifs/IgnatenkoW10}. Fig. \ref{fig:fuzzycommitment} shows the fuzzy commitment scheme. In this scheme, a secret $s\in\{0,1,...,|S|\}$ is independent of biometric data and chosen uniformly at random. So, 

\begin{equation}
Pr\{S=s\}=1/|S|, \quad \forall \,  s\in\{0,1,...,|S|\} \end{equation}

In the enrollment phase, the secret $s$ is fed to the encoder to generate an encoded binary codeword $C^{N}=( \,c_{1},c_{2},...,c_{N}) \,$ such that $c_{i}\in\{0,1\}$, $\forall$ $i=1,2,...,N$. The binary biometric enrollment sequence $X^{N}$ is added ($modulo \;2$) to the generated codeword to produce a sequence $Z^{N}=( \,z_{1},z_{2},...,z_{N}) \,$, such that $z_{i}\in\{0,1\}$, $\forall$ $i=1,2,...,N$. Thus

\begin{equation} 
Z^{N}=C^{N} \oplus X^{N} = e(  \,s) \, \oplus X^{N}
\end{equation}

where, $C^{N}=e(  \,s) \,$ and $e( \,\cdot) \,$ represents the encoding function. The public sequence $Z^{N}$ released to the authentication side is referred to as the \textit{helper data}.

In the authentication, a binary biometric authentication sequence $Y^{N}$ is added ($modulo \;2$) to the helper data $Z^{N}$ received from enrollment. This addition results in the  sequence $R^{N}$.

\begin{equation} 
R^{N}=Z^{N} \oplus Y^{N} = e(  \,s) \, \oplus X^{N} \oplus Y^{N}
\end{equation}

where, $R^{N}=( \,r_{1},r_{2},...,r_{N}) \,$, such that $r_{i}\in\{0,1\}$, $\forall$ $i=1,2,...,N$. Thus, we can consider $R^{N}$ as a codeword $C^{N}$ with some additional noise sequence, i.e., $X^{N} \oplus Y^{N}$. A decoder accepts $R^{N}$ as input to determine the estimate $\hat{s}$ as

\begin{equation}
\hat{s}=d( \,R^{N}) \,=d( \, e( \,s) \, \oplus( \,X^{N} \oplus Y^{N} ) \, ) \,
\end{equation}

where, $d( \,\cdot) \,$ represents the decoding function.

Like a conventional cryptographic commitment scheme, the fuzzy commitment scheme is both concealing and binding: it is infeasible for an attacker to learn the committed value, and also for the committer to decommit a value in more than one way \cite{DBLP:conf/ccs/JuelsW99}. Lafkih et al. \cite{DBLP:journals/iajit/LafkihMGHA16} proposed a security analysis framework based on several scenarios of threats that can affect biometric cryptosystems and applied this analysis on FCS and determined theoretically and practically that cryptosystems based on FCS do not ensure a high level of security or protection of privacy.

A \textit{fuzzy vault scheme} \cite{DBLP:journals/iacr/JuelsS02}, is an order invariant cryptographic construction in the form of an error-tolerant encryption operation where keys consist of sets. It is an extension of the fuzzy commitment scheme. Suppose, Alice possess a secret encryption key \textit{k} and an unordered set \textit{A}. She places \textit{k} in a vault and locks using \textit{A}. If Bob possess another unordered set \textit{B}, then he can unlock the vault to access \textit{k} if and only if \textit{B} largely overlaps \textit{A}.

You et al. \cite{You} defined the fuzzy vault scheme also called as fingerprint vault scheme as follows. In this scheme, initially every extracted minutia is converted into an element $m_i$ in a finite field $\mathbb{F}$. The selected secret key $k$ is encoded to a polynomial $f\left(x\right)\in \mathbb{F}\left[X\right]$, which is evaluated in each $m_i$. All the points $\left(m_{i},f\left(m_{i}\right)\right)$ and some stochastic camouflage points $c_i,d_i$ are selected, such that $d_i \neq f\left(c_i\right)$ are to be stored together as a vault $V$.

The security of this scheme is based on the infeasibility of the polynomial reconstruction problem. The fuzzy vault scheme is an example of the biometric-based key generation. This scheme possesses characteristics that make it suitable for applications that combine biometric authentication and cryptography; the advantages of cryptography (e.g., proven security) and fingerprint-based authentication (e.g., user convenience, non-repudiation) can be utilized in such systems \cite{DBLP:conf/avbpa/UludagPJ05}. Nagar et al. have shown that both performance and security of a fingerprint fuzzy vault can be improved by incorporating minutiae descriptors, which capture orientation and ridge frequency information in a minutia's neighborhood in \cite{DBLP:conf/icpr/NagarNJ08}.

\textit{Shielding function}
Shielding functions are proposed to enhance privacy and prevent misuse of biometric templates in key binding schemes. These function aim to enhance the reliability and reproducibility of the detection and to shield the information (or `entropy') in the secret from the reference data \cite{DBLP:conf/IEEEias/LiWPZ09}. 

\begin{figure}[h]
	\centering
	\includegraphics[width=0.50\textwidth]{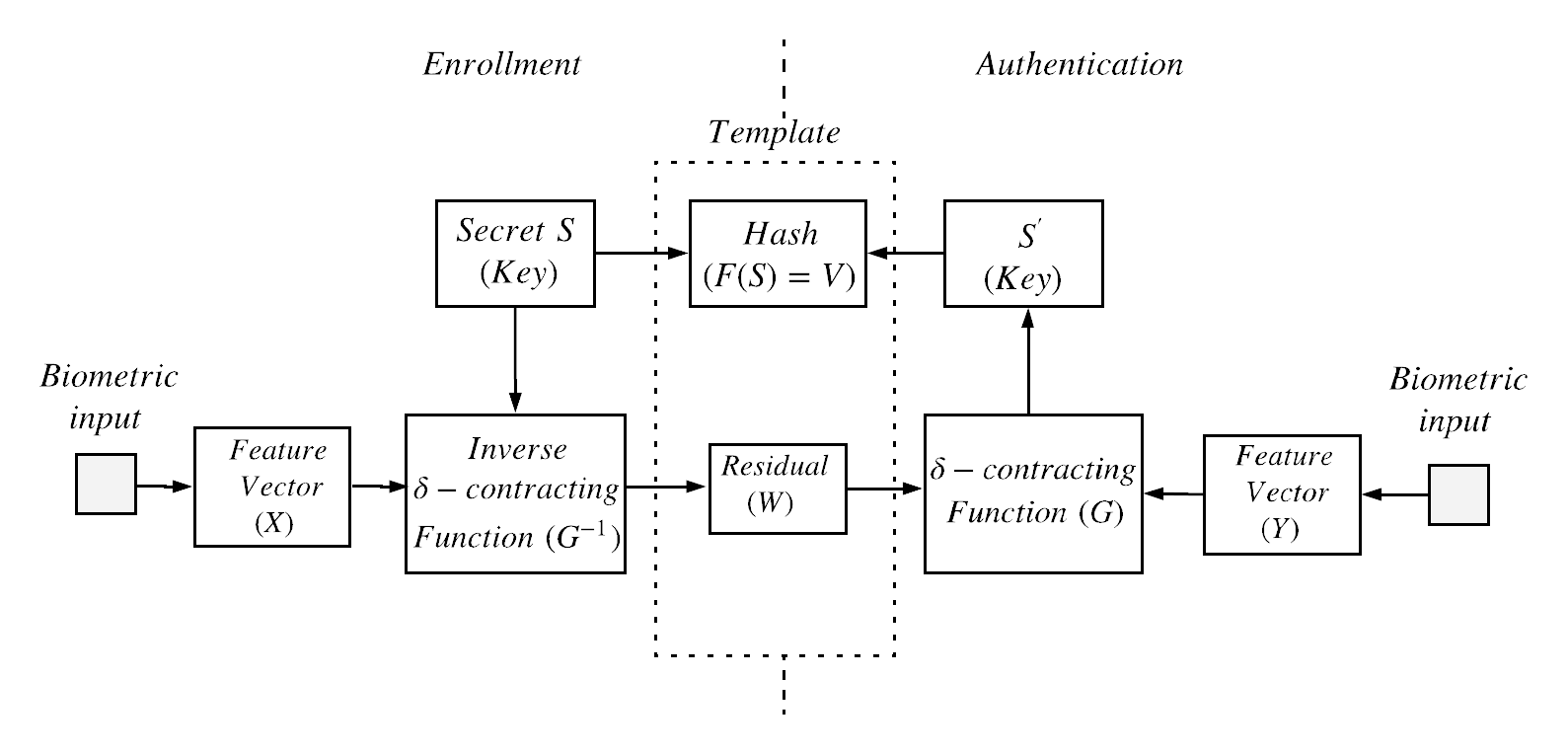}
	\caption{Shielding functions: basic operation mode \cite{DBLP:journals/ejisec/RathgebU11}}
	\label{fig:shieldingfunction}
\end{figure}

Fig. \ref{fig:shieldingfunction} shows the basic operation mode of shielding function. A fixed length, real-valued, noise free 
biometric feature vector $X$ is assumed at the enrollment. A secret $S$ (the key) together with $X$ is applied to an inverse $\delta$-contracting function $G^{-1}$ to generate the helper data $W$, such that $G\left(W,X\right)=S$. A hash of the secret $S$ is stored (as a template) additionally as $F\left(S\right)=V$. The $\delta$-contracting function $G$ that computes a residual for each feature is the core of this scheme. $W$ comprises all residuals that form a correction vector. During authentication phase, a new feature vector $Y$ is used to compute $G\left(W,Y\right)$. If $||X-Y||\leq \delta$, then $G\left(W,Y\right)=S'=S=G\left(W,X\right)$. Finally, the hash value $F\left(S’\right)$ of the reconstructed secret $S’$ is tested against the previously stored one $\left(V\right)$ yielding successful authentication or rejection \cite{DBLP:journals/ejisec/RathgebU11}. 

\subsubsection*{Key-generation scheme}
\begin{figure}[h]
	\centering
	\includegraphics[width=0.50\textwidth]{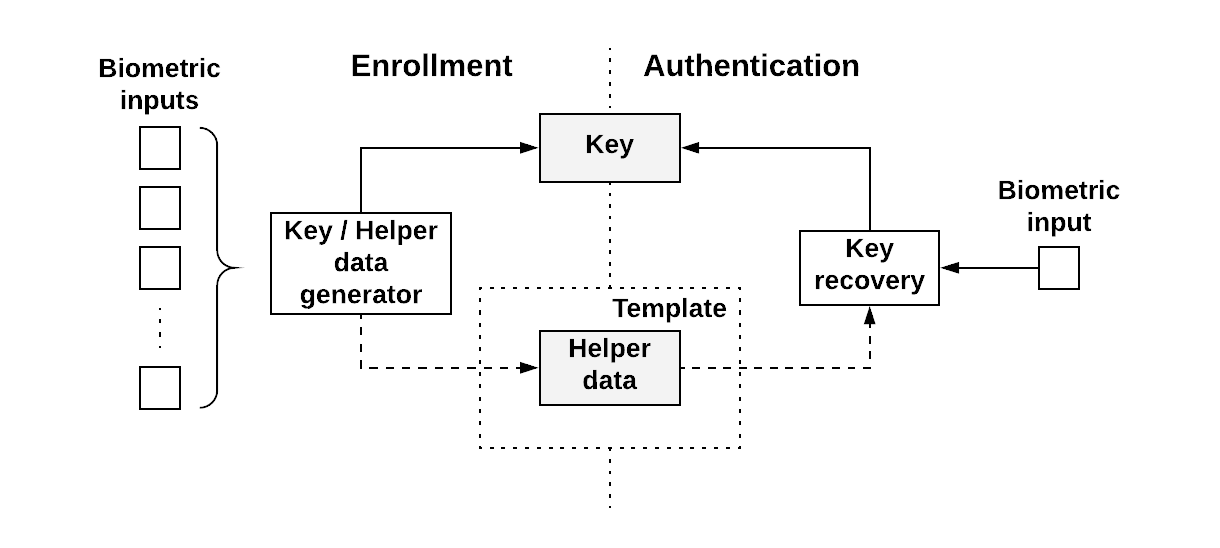}
	\caption{Basic concept of biometric key-generation \cite{DBLP:journals/ejisec/RathgebU11}}
	\label{fig:keygeneration}
\end{figure}

Fig. \ref{fig:keygeneration} shows the basic concept behind the key generation scheme. Key generation techniques extract a cryptographic key from a biometric sample during the enrollment (along with the helper data, if necessary) and the same key has to be extracted using a new biometric sample (and the helper data when available) during authentication \cite{DBLP:journals/eswa/ImamverdiyevTK13}. 

\textit{Quantization}
The idea of quantization is to quantize each component of the
feature vector with a threshold to tolerate the noise \cite{DBLP:journals/jdim/HanWN08}. The best way to protect the user biometrics is to ensure that the information on the template is incomprehensible and irreversible. Hash functions are generally used for such purposes. A hash function is extremely sensitive to input value; the biometric samples from the same user are never exactly the same hence the feature vectors don't match exactly. As a result, the hash function can not be directly used to protect biometric templates. The adaptive non-uniform quantization algorithm maps different noisy feature vectors from one biometric sample to a unique vector and then protect it by cryptographic hash functions, such as MD5.

\textit{Secure sketch and Fuzzy extractor}
Secure sketches and extractors can be viewed as providing fuzzy key storage; they allow recovery of the secret key ($w$ or $R$) from a faulty reading $w'$ of the password $w$ by using some public information ($s$ or $P$) \cite{DBLP:journals/corr/abs-cs-0602007}. The secure sketch scheme consists of two main components, namely, an encoder and a decoder. An encoder (sketch generation algorithm) accepts the original biometric template $w$ as input and gives a sketch (extra information) $s$ as output. The encoder (biometric template reconstruction algorithm) accepts a live biometric template $w'$ and the sketch $s$ as input and generates $w$ only if $w$ and $w'$ are sufficiently close to each other according to some similarity measure.

\begin{figure}[h]
	\centering
	\includegraphics[width=0.50\textwidth]{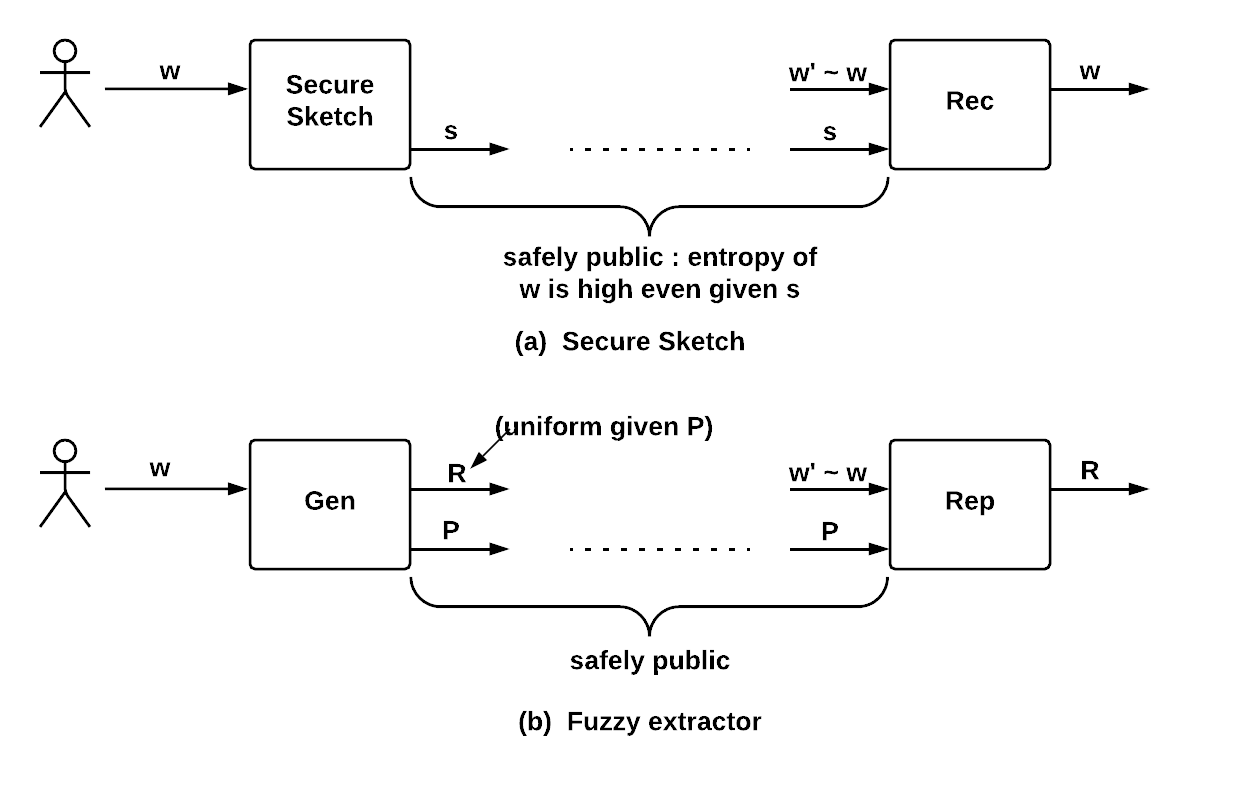}
	\centering
	\caption{(a) Secure sketch : On input $w$, a procedure outputs a sketch $s$. Then, given $s$ and a value $w'$ close to $w$, it is possible to recover $w$. (b) Fuzzy extractor : It extracts a uniformly random string $R$ from its input $w$ in a noise-tolerant way. Noise tolerance means that if the input changes to some $w'$ but remains close, the string $R$ can be reproduced exactly \cite{DBLP:journals/corr/abs-cs-0602007}.}
	\label{fig:securesketch}
\end{figure}

The fuzzy extractors allow one to extract some randomness $R$ (using generation procedure $Gen$) from $w$ and then successfully reproduce $R$ (using reproduction procedure $Rep$) from any string $w'$ that is close to $w$ \cite{DBLP:journals/corr/abs-cs-0602007}. For an efficient fuzzy extractor, the procedures $Gen$ and $Rep$ must run in expected polynomial time.

\subsubsection*{Key-release scheme}
In a \textit{key release mode}, biometrics plays a predetermined role in a cryptosystem, and the key would be released to users only if biometric matching is successful \cite{4743423}. The user authentication process is completely decoupled from the key release operation.

\begin{figure}[h]
	\centering
	\includegraphics[width=0.35\textwidth]{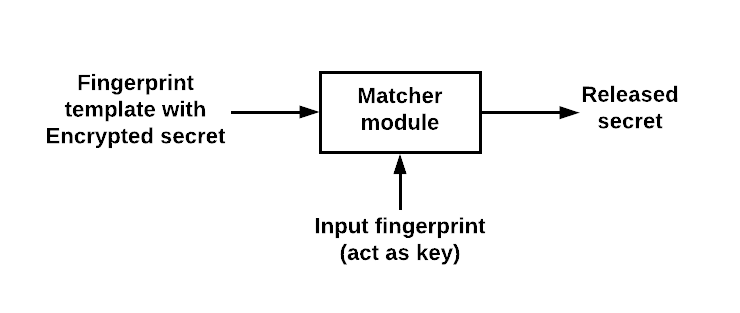}
	\caption{Key release scheme : In the fingerprint-based authentication, a cryptographic key is the ``secret'' and fingerprint is the ``key'' and the cryptographic key is released upon a successful authentication. \cite{Maltoni:2009:HFR:1557624}.}
	\label{fig:keyrelease}
\end{figure}

Fig. \ref{fig:keyrelease} shows the key release scheme of biometric template protection. In such systems, a cryptographic key is stored as part of a user’s database record, together with the user name, biometric template, access privileges etc., that is only released upon a successful biometric authentication \cite{DBLP:journals/pieee/UludagPPJ04}. The system using key release scheme, uses local storage of the biometric system, such as smart card, which raises issue about stealing user biometrics. Moreover, these systems are also vulnerable to Trojan horse attack at the decision module. Such an attack will bypass all system components and inject 1-bit accept/reject decision code into the system decision module.

\subsubsection{Feature transformations}
In the feature transformation based approach, a transformation function is applied to the template, and the output transformed template is stored in the database. There are various transformation functions available in the literature, e.g. cancelable biometric \cite{DBLP:conf/icpr/RathaCBC06}, biohashing \cite{DBLP:conf/biosig/TopcuEKY13}, biometric salting \cite{Jain:2008:BTS:1376536.1387883}, MRP (Multispace Random Projection) \cite{DBLP:journals/tsmc/TeohY07} etc. Some of them are discussed below.

\subsubsection*{Cancelable biometrics}
The non-invertible transformation schemes, such as the \textit{cancelable biometrics} apply a {\em one-way function}\footnote{a function $f\left(x\right)$ is said to be a one-way function if it is hard to invert, i.e., given a value $y$, it is computationally infeasible to find some input $x$ such that $f\left(x\right)=y$} to the template rendering it computationally hard for the adversary to invert the template. The cancelable biometrics on large fingerprint database was demonstrated by Ratha et al. in \cite{DBLP:conf/icpr/RathaCBC06}. Their results proved that a large number of cancelable fingerprint transformations could be constructed without losing much accuracy of the biometric system. Cancelable biometrics consist of intentional, repeatable distortions of biometric signals based on transforms which provide a comparison of biometric templates in the transformed domain \cite{Ratha}. Jin et al. \cite{DBLP:journals/pr/JinTGT16} proposed an error-correction code (ECC) free key binding scheme along with cancelable transform for minutia-based fingerprint biometrics. The analysis done by Nagar et al. demonstrates that cancelable fingerprint templates are vulnerable to intrusion and (template) linkage attacks because it is relatively easy to obtain a pre-image of the transformed template \cite{DBLP:conf/mediaforensics/NagarNJ10}.

\textit{Biometric salting} is a template protection approach in which the biometric features are transformed using a function defined by a user-specific key or password \cite{Jain:2008:BTS:1376536.1387883}. Biometric salting usually denotes transformations of biometric templates, which are chosen to be invertible \cite{DBLP:journals/ejisec/RathgebU11}. The adversary can recover the original template if he accesses the key and the template. Therefore, security of salting-based approach depends on the secrecy of the key and the template. The key must be secret in private domain, and must be used for every authentication attempt.

\textit{Biohashing}, applied to fingerprint biometric, is a two factor authentication approach that combines fingerprint feature with a user specified key/token and generates a unique compact code per person \cite{DBLP:conf/biosig/TopcuEKY13}. It is a product of the biometric feature of an individual (such as a fingerprint) with a tokenized pseudo-random number that generates a set of user-specific secret code called as \textit{biohash}. 

\begin{figure}[h]
	\centering
	\includegraphics[width=0.50\textwidth]{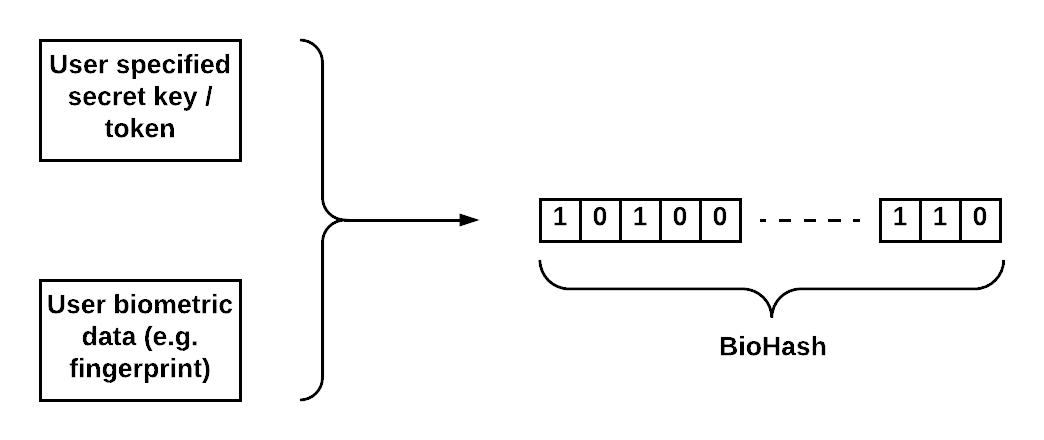}
	\caption{BioHashing : Two factor authentication - secret key and biometric data \cite{DBLP:conf/biosig/TopcuEKY13}}
	\label{fig:biohash}
\end{figure}

Fig. \ref{fig:biohash} shows the generation of a biohash with the help of a user's secret information (e.g. secret key or token) and his biometric data, such as fingerprint. An integrated wavelet and Fourier–Mellin transform (WFMT) is used to create the biometric features since, in WFMT framework, wavelet transform preserves the local edges and noise reduction in the low-frequency domain (high energy compacted) after the image decomposition. This renders the fingerprint images less sensitive to shape distortion. The analysis done by Nagar et al. in \cite{DBLP:conf/mediaforensics/NagarNJ10} indicates that biohashing is vulnerable to intrusion and (template) linkage attacks because it is relatively easy to obtain a close approximation of the original template.

\textit{Multispace Random Projection (MRP)} :
Teoh et al. \cite{DBLP:journals/tsmc/TeohY07} proposed a generic cancelable biometrics formulation (coined as MRPs), which is a two-factor cancelable formulation, where the biometric data is distorted in a revocable but non-reversible manner by first transforming the raw biometric data into a fixed-length feature vector. Subsequently, the feature vector is projected onto a sequence of random subspaces that were derived from a user-specific pseudo-random number (PRN). The raw biometric data is transformed into a fixed-length feature vector, and then the non-invertible random subspace
projection is performed on it to carryout MRP. A token or a centered database stores the user-specific seeded pseudo-random number generator. The independent, zero-mean, and unit-variance Gaussian-distributed random bases generated from PRN generators constitutes the random subspace. MRP can be revoked through the pseudo-random numbers replacement, so that a new template can be generated instantly, and its non-invertible property helps to preserve the privacy of the user biometrics data \cite{DBLP:journals/tsmc/TeohY07}.

\textit{Robust hashing} :
Robust hashing is a secure biometric based authentication scheme
which employs a user-dependant one-way transformation
combined with a secure hashing algorithm \cite{DBLP:conf/mmsec/SutcuSM05}. The robust hashing function ensures the security and privacy of biometric data, since it is designed as a sum of properly weighted and shifted Gaussian functions. Preliminary results show that proposed scheme offers a simple and practical solution to one of the privacy and security weakness of biometrics-based  authentication systems namely, template security \cite{DBLP:conf/mmsec/SutcuSM05}.

\textit{BioPhasoring} :
BioPhasor, a generic cancelable biometrics formulation, is a set of binary codes based on iterated mixing between the user-specific tokenized pseudo-random number and the biometric feature that enables straightforward revocation of biometric template via token replacement \cite{DBLP:conf/icarcv/TeohN06}. Whenever a genuine token is used, BioPhasor offers high accuracy during verification. As it is a general framework, different biometrics, such as fingerprint, face, and iris, can be secured using BioPhasor approach. Due to the highly stable binary representation in BioPhasor, it can be easily incorporated into biometric encryption techniques.

\textit{Non-invertible transforms} :
There is a significant intra-class variation in the fingerprint representations; multiple acquisitions of the same finger lead to different number of minutia as well as their position $\left(x,y\right)$ and orientation $\left(\theta\right)$, due to which it becomes difficult to match fingerprints in encrypted domain \cite{DBLP:conf/wifs/NagarJ09}. A minutia based fingerprint template $T$ consists of $n$ minutia points such that, $T=\left\{\left(x_1,y_1,\theta_1\right),\left(x_2,y_2,\theta_2\right),\cdots,\left(x_n,y_n,\theta_n\right)\right\}$, where $\left(x,y\right)$ represents the position of each minutia and $\left(\theta\right)$ is its orientation. The transformation function, $\phi\left(\cdot\right)$, transforms $T$ into a new set of $n$ minutia, i.e., $\phi\left(T\right)=\left\{\left(x_1',y_1',\theta_1'\right),\left(x_2',y_2',\theta_2'\right),\cdots,\left(x_n',y_n',\theta_n'\right)\right\}$. Thus, a measure of non-invertibility estimates the difficulty in obtaining $T$ given $\phi\left(T\right)$ \cite{DBLP:conf/wifs/NagarJ09}.

\textit{Cartesian and polar transformation} :
In the Cartesian transformation, the minutiae positions are
measured in rectangular coordinates with reference to the position of the singular point, where the x-axis is aligned with the orientation of the singular point \cite{Ratha:2007:GCF:1263142.1263435}. The coordinate system is divided into fixed sized cells numbered in a fixed sequence. The Cartesian transformation changes the initial cell positions (with rotation in multiple of $90^{\circ}$ after transposition. In polar transformation, the minutiae positions are measured in polar coordinates with reference to the core position \cite{Ratha:2007:GCF:1263142.1263435}. The orientation of the core in the fingerprint is used as reference to measure the angle. In this case, the coordinate space is divided into polar sectors ($L$ levels and $S$ angles) numbered in sequence. The transformation changes the initial sector positions along with changing the minutia angles in accordance to the difference in the sector positions before and after transformation. Ahmad et al. \cite{Ahmad} proposed a local feature-based, and Cartesian and polar transformation-based cancelable fingerprint template scheme.

\textit{Surface folding} :
In a surface folding transformation, both the position and the orientation of the minutia are changed by a mapping function  \cite{Maltoni:2009:HFR:1557624}. Conceptually, in this transformation the minutia are embedded in a sheet and then the sheet is crumpled.

\begin{figure}[h]
	\centering
	\includegraphics[width=0.50\textwidth]{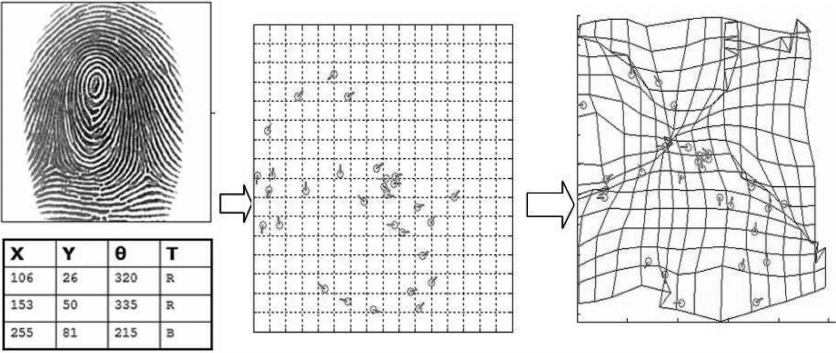}
	\caption{Surface folding transformation \cite{Maltoni:2009:HFR:1557624}}
	\label{fig:surfacefolding}
\end{figure}

Fig. \ref{fig:surfacefolding} shows the surface folding transformation. It is observed that the transform is smooth locally, but not globally. Also, the transform has ``folds'', i.e., multiple minutia location in the original space map to same location in the transformed space. This property provides the desired non-invertibility as it creates ambiguity while reversing the transform. However, due to fairly low ``degree of folding'' (after the transformation only 8\% minutia have their neighbors perturbed) the non-invertibility provided by functional surface folding transform is not strong in spite of high matching accuracy.

\textit{Block permutations} :
Block permutation is a non-invertible transformation technique for cancelable biometric system. This involves the process of splitting the entire image output by the biometric sensor, into several smaller blocks which are shuffled based on a key value, such that the entire input image becomes chaotic \cite{PUNITHAVATHI20178}. The application of this technique to fingerprint biometric system involves the division of the original fingerprint minutia image into several blocks. A user secret key is used to shuffle these blocks in random order so that the real minutia positions are not revealed to the adversary.

\begin{figure}[h]
	\centering
	\includegraphics[width=0.50\textwidth]{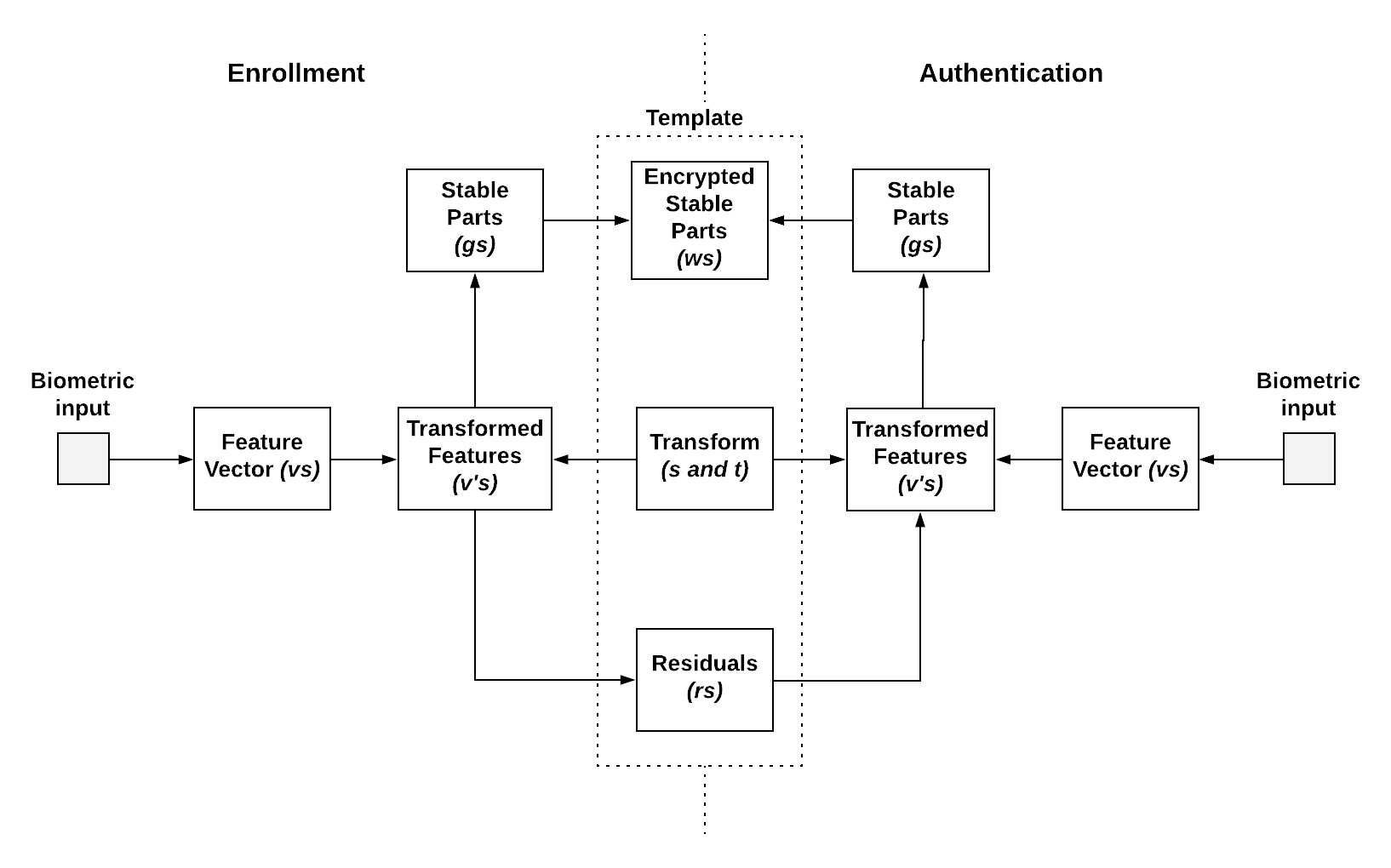}
	\caption{Biotokens: basic operation mode \cite{DBLP:journals/ejisec/RathgebU11}}
	\label{fig:biotokens}
\end{figure}

\textit{Biotokens} :
Fig. \ref{fig:biotokens} shows the basic operation mode of biotokens. Initially, the scaling and translation transformation is applied to each measured biometric feature $v$ to produce $v'=\left(v-t\right)\cdot s$. The transformed features are then split into a stable part $g$ (an integer) and an unstable part $r$. The first part of the secure biometric template, $w$, is a one-way transform of $g$. The unencoded $r$ obscured by the transform along with $s$ and $t$ form the second part of the template.

In authentication, features are transformed by applying $s$ and $t$ onto a residual region defined by $r$ and the unencrypted $r$ is used to compute the local distance within a ``window'', which is
referred to as \textit{robust distance measure}, to provide a perfect match of $w$ \cite{DBLP:journals/ejisec/RathgebU11}. Scheirer et al. introduced \textit{bipartite biotoken} a secure template construct for cryptographic transactions supporting biometrics that, when re-encoded, can release session specific data in \cite{DBLP:conf/icb/ScheirerB09}. The authors showed that bipartite biotokens offer a convenient enhancement to keys and passwords, allowing for tighter auditing and non-repudiation, as well as protection from phishing and man-in-the-middle attacks \cite{biotokens12}. 

\textit{BioConvolving} :
Maiorana et al. introduced a novel noninvertible transform-based approach, namely, BioConvolving, which provides both protection and renewability for any biometric template \cite{Maiorana:2010:CTS:1820799.1820808}. The transform can be expressed in terms of a set of discrete sequences related to the temporal, spatial, or spectral behavior of the considered biometrics. The BioConvolving approach is characterized by three types of transformations which, when applied to an original biometric template, $R_F$ with $F$ sequences $r_{\left(i\right)} \left[n\right]$, of length $N$, can generate a secure template $T_F$ composed by $F$ functions $f_{\left(i\right)} \left[n\right], i=1,2,...,F$, from which retrieving the original data is as hard as randomly guessing them \cite{bioconvolving}. The security of the BioConvolving approach depends on the fact that, if an impostor gains access to the stored information, he has to resolve a blind deconvolution problem to retrieve the original template \cite{DBLP:journals/eswa/NanniMLC10}. The advantage of BioConvolving scheme is that it is not feasible for the adversary to retrieve the original template even if he succeeds in stealing more than one transformed template. However, the primary drawback of the BioConvolving cancelable transformation system was the length of its transformed template, which does not have the same length as the original one \cite{bioconvolving2}.

\textit{Fingerprint shell} :
Moujahdi et al. exploits the information provided by the
extracted minutiae to construct a new representation based on special spiral curves (that will be stored in the database as template). The representation can be used for the recognition task instead of the traditional minutiae-based representation in \cite{DBLP:journals/prl/MoujahdiBGR14}. Initially, the minutia and singular points (such as, core and delta) are extracted from the fingerprint image. The distance between each minutia and every core and delta point is calculated. These distances are arranged in ascending order. These distances form the hypotenuses of several right angle triangles constructed in the next step. The algorithm keeps only the spiral curves thus formed for matching in the authentication phase. The authors demonstrated that the fingerprint shell is invulnerable to brute force attack and zero-effort attack. The limitations of this approach include a drastic variation in the constructed curve due to the addition of spurious minutia or missing of majority of minutia. Due to this limitation, the fingerprint shell is not efficient enough to be used in biometric acquisition in all scenarios.

\subsubsection{Template image transformation}
While biometric techniques have inherent advantages over
traditional personal identification techniques, the problem of
ensuring the security and integrity of the biometric data is critical, e.g. if a person’s biometric data (such as, her fingerprint image) is stolen, it is not possible to replace it unlike replacing a stolen credit card, ID, or password \cite{DBLP:journals/pami/JainU03}. Encryption, steganography, watermarking and visual cryptography are possible solutions to enhance the security and integrity of biometric data.

\subsubsection*{Steganography}
Steganography communicates secret information by hiding it in multimedia carrier such as image/video/audio files \cite{DBLP:journals/mta/KaurK16}. A novel and efficient information security system to protect biometric signatures from any unauthorized access which employs orthogonal coding scheme, encoded steganography, and nonlinear encryption scheme involving joint transform correlation is proposed in \cite{ISLAM20154026}. The technique employs orthogonal coding scheme to encode multiple biometric information and multiplex together which makes it almost impossible to decode even a part of the secret information without authorization \cite{ISLAM20154026}.

\subsubsection*{Watermarking}
Ratha et al. \cite{DBLP:conf/mm/RathaCB00} described a robust data-hiding algorithm in the wavelet-compressed domain for fingerprint images which is simple and can be easily implemented in hardware. The approach provides security against replay attack for an on-line fingerprint authentication system. In this method, a different verification string is issued by the service provider (server) for each transaction. The verification string is combined with the fingerprint image, and then the new image is transmitted over the insecure communication channel. The server on receiving this image back can decompress it to verify the presence of correct string embedded in it. Thus the replay of stored images to the matcher is prevented. Jain et al. introduced two applications of an amplitude modulation-based watermarking (hiding a user’s biometric data in a variety of images) method which has the ability to increase the security of both the hidden biometric data (e.g., eigen-face coefficients) and host images (e.g., fingerprints) \cite{DBLP:journals/pami/JainU03}. The authors claimed that a high accuracy in verification on watermarked (e.g., fingerprint) images is guaranteed by the feature analysis of the host images. 

\subsubsection*{Visual Cryptography Scheme (VCS)} 
Naor et al. introduced a cryptographic scheme, visual cryptography, which is perfectly secure, very easy to implement and can decode concealed images without any cryptographic computations in \cite{10.1007/BFb0053419}. VCS is a cryptographic technique that allows for the encryption of visual information such that the decryption can be performed using the human visual system \cite{DBLP:journals/tifs/RossO11}. The basic VCS scheme is denoted as $\left(k,n\right)$ VCS, and read as $k-out-of-n$ VCS. Suppose $T$ is the original binary image, the scheme encrypts it in $n$ images such that, 

\begin{equation}
T=S_{h_1} \oplus S_{h_2} \oplus S_{h_3} \cdots \oplus S_{h_k}
\end{equation}

where $\oplus$ is a Boolean operation and $S_{h_i}$ is an image that appears as white noise, such that, $h_i\in (1,2,3, \cdots ,k), k \leq n$ and $n$ specifies the number of noisy images. The encryption is undertaken in such a way that $k$ or more out of the $n$ generated images are necessary for reconstructing the original image $T$ and it is difficult to decipher the secret image
using individual $S_{h_i}$’s \cite{DBLP:journals/tifs/RossO11}.

\subsection{Secure Multiparty Computation (SMPC)}
The Secure Multiparty Computation technique is useful in a system composed of $n$ entities, namely $e_{1},e_{2},...,e_{n}$. All entities jointly compute a public function $g$ using individual secret data, $s_{1},s_{2},...,s_{n}$, keeping their private inputs undisclosed from each other.

\subsubsection*{Homomorphic Encryption (HE)} 
Gomez-Barrero et al. proposed a new efficient biometric template protection scheme based on homomorphic probabilistic encryption for fixed-length templates, where only encrypted data is processed \cite{DBLP:conf/cvpr/Gomez-BarreroFG16}. Homomorphic Encryption schemes require no auxiliary data (which can be exploited
in order to obtain information about the hidden biometric data, thus violating the privacy of the subject) and allow for computations to be performed on ciphertexts, which generate encrypted results whose corresponding plaintexts match the results of the operations carried out on the original plaintext \cite{DBLP:conf/cvpr/Gomez-BarreroFG16}. Most HE schemes rely on asymmetric cryptography, and the homomorphic property holds under encryption with the public key of one of the parties involved in the protocol \cite{DBLP:journals/spm/BarniDL15}. 

In an honest-but-curious adversary model \cite{Goldreich:2004:FCV:975541}, where both parties, client and server, follow the established protocols but may try to learn additional information about the other side template, the server must process the client’s biometric data without extracting any information from it, and at the same time, the server must protect the information stored in the database \cite{DBLP:journals/spm/BarniDL15}. To achieve this, the Paillier homomorphic probabilistic encryption scheme \cite{DBLP:conf/eurocrypt/Paillier99} is used, which is based on the decisional composite residuosity assumption: given a composite $n$ and an integer $z$, it is hard to decide whether $z$ is an $n$-residue modulo $n^{2}$. 

\subsubsection*{Oblivious Transfer (OT)}
Oblivious transfer is a secure two-party computation (STPC) protocol which enables one party, say the sender (e.g. a server) $S$, to send one out of $n$ messages $(x_{1},x_{2},...,x_{n})$ to a receiver (say, a client) $R$. The receiver initially chooses any of the $n$ messages. The communication completes without the receiver revealing any information about the chosen $x_{i}$ to the sender and also the information about the remaining $n-1$ messages is kept hidden from the receiver. If we consider the elements to be the stored (encrypted) biometric templates, we see that OT essentially allows one to search in the database, without revealing which item (i.e., biometric template) is selected for the matching process \cite{DBLP:journals/scn/PagninM17}. OT alone cannot prevent some
template recovery attacks, since the best known strategy is
based solely on the value returned by the (Biometric Authentication System) BAS (essentially the acceptance/rejection message) which is not affected by the OT technique \cite{DBLP:journals/scn/PagninM17}. Hamming distance together with oblivious transfers is one of the most elegant tools used in biometric authentication systems \cite{DBLP:journals/scn/KirazGK15}.

\subsubsection*{Garbled Circuits (GC)} 
Garbled circuits are a cryptographic technique (combining OT and SMPC between two entities) that enables two parties to compute a function (represented as a binary circuit) and learn only the output of the function and nothing else (e.g., the other party’s input) and thus is quite relevant for achieving a privacy-preserving matching process in biometric authentication \cite{DBLP:journals/scn/PagninM17}. Given that the complexity depends on the number of gates composing the circuit, GCs are suited for operations such as sums and comparisons, for which the number of gates depends linearly on the input bit length \cite{DBLP:journals/spm/BarniDL15}.

Dodis et al. coined the terms \textit{secure sketch} and \textit{fuzzy extractor} in the context of key generation from biometric data \cite{DBLP:journals/corr/abs-cs-0602007}. The general idea of \textit{fuzzy extractor} is based on a two-step process, where an extraction function first transforms any sufficiently random fuzzy secret into an almost uniform random private string, and outputs some public information, which is used in the regeneration step to reconstitute the exact same private string from a close enough approximation of the original fuzzy secret \cite{DBLP:conf/ccs/Boyen04}.

Umut Uludag et al. \cite{DBLP:journals/pieee/UludagPPJ04} have presented various methods that monolithically bind a cryptographic key with the user's stored biometric template in such a way that the key cannot be revealed without a successful biometric authentication. In \cite{DBLP:journals/eswa/ImamverdiyevTK13}, the authors have proposed an effective biometric cryptosystem construction for biometric template protection using a number of discretized fingerprint texture descriptors and appropriate error correcting codes (ECC). Jo et al. have proposed a biometric digital key generation/extraction mechanisms for cryptographic secure communication \cite{DBLP:conf/faw/JoSL07}. Arjona et al. provided the implementation of a fingerprint biometric cryptosystem in FPGA \cite{10261/122760}. 

A novel algorithm for fingerprint encryption which transforms fingerprint minutia and performs matching in the transformed form was proposed in \cite{DBLP:journals/prl/ChenC11}. This algorithm constructs a circular region around each minutia and encrypts these regions before storing as ``canceled template'' in the database. Considering their results, the authors claim to have achieved improvement in accuracy compared to fuzzy vault system. Benjamin Tams investigated the implementations of biometric cryptosystems for protecting fingerprint templates and concluded that a single fingerprint is insufficient to provide a secure biometric cryptosystem in \cite{DBLP:journals/corr/abs-1304-7386}. 

The biometric data can be secured using encryption techniques, watermarking, and steganography. The use of an amplitude modulation-based watermarking method to hide a user’s biometric data in a variety of images so as to increase the security of both the hidden biometric data and host images (e.g., fingerprints) was proposed in \cite{DBLP:journals/pami/JainU03}. The hashing technique can also be used to secure biometric data. A method hashing fingerprint minutia information, wherein only hashed data is transmitted and stored in the server database, and it is not possible to restore fingerprint minutia locations using hashed data in \cite{DBLP:conf/icapr/TulyakovFG05}. 

\section{Issues and challenges}\label{issues}

The fingerprint biometric system is widely used in border control, forensics, criminal identification, access control, computer logins, e-commerce, welfare disbursements, missing children identification, id-cards, passports, user authentication on mobile devices, time and attendance monitoring systems \cite{DBLP:journals/pr/UnarSA14}. Due to the wide acceptance of the biometric system as a reliable means for user identification and authorization, it becomes a need to address various challenges for improving the system performance and enhance the security of such systems. Jain et al. have categorized the fundamental barriers in biometrics into four main categories, namely, accuracy, scale, security, and privacy in \cite{DBLP:conf/icpr/JainPPHR04}. The accuracy of a biometric system is highly influenced by false-match and false-nonmatch errors in making the correct decision. The scale barrier poses a question about the effect of the number of enrolled users in making a correct decision. The security of the biometric system against possible attacks and the privacy of user data is of paramount importance.

The future of biometric system along with various research opportunities are discussed in \cite{PatoBiometricRC}. The authors have addressed opportunities in modality-related research, information security research, testing and evaluation research, systems-level statistical engineering research, research on scale and social science. The privacy issues related to biometric system are discussed in \cite{DBLP:journals/ieeesp/PrabhakarPJ03}. The privacy issues are mainly related to user data (i.e. biometrics). Privacy concerns arise when biometric data are used for secondary purposes, invoking function creep, data matching, aggregation, surveillance and profiling \cite{DBLP:conf/ic3/PanigrahyJKJ09}. To address the privacy issue of an individual's biometrics, a fingerprint authentication system for the privacy protection of the fingerprint template stored in a database is introduced in \cite{DBLP:journals/spl/LiK11}. In this case, the identity of an individual is hidden in a thinned fingerprint template, which is stored in an online database and retrieved in authentication phase. Some people do not believe in the biometric system due to privacy concerns. Hence, the research community is facing a challenge to develop a biometric system which takes care of not only the security but also the privacy of the user data.

The World Bank estimates that more than $1.5$ billion people worldwide do not officially exist \cite{STORISTEANU20165}. Biometric system can be the possible solution to this global issue of rendering the resources and services of the government accessible to the people.

Akhtar et al. have performed a comprehensive study on biometric systems and tried to answer some of key questions related to biometrics in \cite{7948983}. The authors have categorised the queries into various sections, such as the current status of biometrics, current issues and challenges of biometrics, hot topics in biometrics, security of biometrics and future of biometrics. 

\section{Open problems and Opportunities}\label{problems}
The biometric systems are improving every day making our lives better and secure. In this Section, we identify present limitations, open problems, and future research opportunities in biometric systems. Table \ref{issues} shows the topics in biometrics that still need attention from industry and academia.

\begin{table}[h]
	\centering
	\caption{Opportunities in fingerprint biometric system}
	\label{issues}
	\begin{tabular}{|l|l|}
		\hline
		\multicolumn{1}{|c|}{\textbf{Topic}}                                         & \multicolumn{1}{c|}{\textbf{Issues / research  opportunities}}                                                                                \\ \hline
		\begin{tabular}[c]{@{}l@{}}Smartphone \\ biometrics\end{tabular}             & \begin{tabular}[c]{@{}l@{}}Improving matching speed and addressing\\ template security on smartphone\end{tabular}                             \\ \hline
		\begin{tabular}[c]{@{}l@{}}Big data \\ techniques\end{tabular}               & Securing database against template leakage                                                                                                    \\ \hline
		\begin{tabular}[c]{@{}l@{}}Biometric sensor \\ interoperability\end{tabular} & Sensor technology enhancement                                                                                                                 \\ \hline
		\begin{tabular}[c]{@{}l@{}}Wearable \\ biometrics\end{tabular}               & Addressing privacy issues                                                                                                                     \\ \hline
		\begin{tabular}[c]{@{}l@{}}Performance \\ improvement\end{tabular}           & \begin{tabular}[c]{@{}l@{}}Improving matching speed, narrowing \\ zero-effort attacks\end{tabular}                                            \\ \hline
		\begin{tabular}[c]{@{}l@{}}Sensor technology \\ enhancement\end{tabular}     & \begin{tabular}[c]{@{}l@{}}Improving the sensor speed and making them\\ invulnerable to presentation attacks\end{tabular}                     \\ \hline
		\begin{tabular}[c]{@{}l@{}}Poor quality image\\ recognition\end{tabular}     & \begin{tabular}[c]{@{}l@{}}Improving the digital image processing \\ techniques\end{tabular}                                                  \\ \hline
		\begin{tabular}[c]{@{}l@{}}Biometrics on \\ smartcard\end{tabular}           & \begin{tabular}[c]{@{}l@{}}Issues related to tampering attack and \\ a side-channel attack on this system\end{tabular}                        \\ \hline
		\begin{tabular}[c]{@{}l@{}}Fingerprint \\ indexing\end{tabular}              & Designing techniques for database indexing                                                                                                    \\ \hline
		\begin{tabular}[c]{@{}l@{}}Biometric \\ cryptosystems\end{tabular}           & \begin{tabular}[c]{@{}l@{}}Performance enhancement using hardware based\\ multimodal biometric cryptosystem\end{tabular}                      \\ \hline
		\begin{tabular}[c]{@{}l@{}}Touchless \\ biometric\end{tabular}               & Touchless biometric system using multiple traits                                                                                              \\ \hline
		\begin{tabular}[c]{@{}l@{}}Cancelable \\ biometrics\end{tabular}             & \begin{tabular}[c]{@{}l@{}}Cancelable template generation for multimodal\\ biometric systems\end{tabular}                                     \\ \hline
		\begin{tabular}[c]{@{}l@{}}2D-3D \\ biometrics\end{tabular}                  & \begin{tabular}[c]{@{}l@{}}Develop improved image processing techniques\\ to transform 2D biometrics into 3D\end{tabular}                     \\ \hline
		\begin{tabular}[c]{@{}l@{}}Multimodal \\ biometric system\end{tabular}       & \begin{tabular}[c]{@{}l@{}}Issues in design, testing, accuracy, \\ and performance of such systems\end{tabular}                               \\ \hline
		\begin{tabular}[c]{@{}l@{}}Biometrics as \\ digital signature\end{tabular}   & \begin{tabular}[c]{@{}l@{}}Mitigating the hill-climbing attack and issues\\  related to practicable biometric digital  signature\end{tabular} \\ \hline
	\end{tabular}
\end{table}

\textit{Smartphone biometrics} is the use of biometrics to secure a smartphone from unauthorized access and theft. At present more than 95\% of world population is using a smartphone. The use of fingerprint, iris, and face as owner identification is already available with some smartphones. The time required for authentication using biometrics is relatively high, and with multi-factor authentication, such as password and biometrics, things are getting worse for the user. A hardware-level matching can improve the execution time of such systems.

\textit{Big data techniques} for biometrics is required for an extensive template database such as Aadhaar database or a global passport validation with user biometrics. The big data techniques work at data storage, data analytics, efficient searching, and data privacy. These techniques can be a possible solution to a sufficiently sizeable biometric template database. 

\textit{Biometric sensor interoperability} is the adaption of a biometric system to the raw biometrics obtained from the sensors of another biometric system. The biometric systems are tuned to accept the biometrics generated from a single sensor. Hence, they show an abnormal response to the biometrics not produced from its sensors. This issue can be addressed at a level that a capacitive sensor accepts the biometrics obtained from any other capacitive sensor.

\textit{Wearable biometrics} refers to person identification technology incorporated into items of garments and accessories that can read, record, and compare individual’s biometric traits
such as heart rate, respiratory rate, or any physical activity \cite{Blasco:2016:SWB:2988524.2968215}. The devices embedded into daily wearables such as a watch, jewelry, shoes, spectacles, handbags, clothing, etc. provide continuous interaction with the biometric system without user knowledge and intervention. Wearable biometric is the future of user identification. An improvement in sensor technology, such as sound sensor, location sensor, light sensor, temperature sensor etc., and wireless communication protocol is required for wearable biometrics. Biometric signals should be distinguishable, permanent, collectable, and difficult to
imitate \cite{Blasco:2016:SWB:2988524.2968215}. 

\textit{Performance improvement} in biometric system deals with the matching time during the authentication phase. The chances of zero-effort attack should be narrowed down, and at the same time, the user identification and authentication must happen in no time since the biometric submission. To address this problem, we need to revisit the system in its entirety and preserve the security of the system. The biometric input device can be designed to capture multiple fingerprint images and keep the best quality image to improve matching accuracy and minimize the possibility of zero-effort attack. Based on the images captured in the enrollment phase, the predefined threshold should be automatically adjusted for every user, and this user-specific threshold should be used in authentication phase. 

\textit{Sensor technology enhancement} needs to be addressed for detecting and preventing all kinds of fake biometric submission incidents at the sensor level itself. The sensor should be able to sense the physical presence of a user when the biometrics is submitted. An improvement in sensor technology can reduce the probabilities of false acceptance and false rejection.

\textit{Poor quality image recognition} is required since the human and environmental factors affect the way sensor captures the biometrics of an individual. The image enhancement module should be able to make a poorly captured image recognizable by the matcher against the stored template. The image processing techniques can solve this problem.

\textit{Biometrics on smartcard} is already in use, such as Match-on-card and Match-in-sensor wherein the biometric is stored in the smartcard and a one-to-one matching is performed in the authentication mode. The template security and performance of this system have to be evaluated and improved. 

\textit{Fingerprint indexing} is the organization of fingerprint templates stored in the database. The linear search performed on entire database results in a large search space and thus requires \BigO{n} comparisons in the worst case for \textit{n} templates in the database. A relatively less number of worst case comparisons can reduce the search time and ultimately improve the overall performance of the system.

\textit{Biometric cryptosystems} based on a single finger cannot provide significant security since it is susceptible to brute force attack and zero-effort attack, i.e., false acceptance. Also, the biometric cryptosystem is found to be relatively slow as compared to the cancelable biometrics. A hardware-based biometric cryptosystem with multi-modal biometric can address this problem.

\textit{Touchless biometric} system senses the biometrics of an individual with physical contact with the sensor. A palm print based touchless biometric system is available as a commercial product. A multimodal touchless biometric system can efficiently enhance the security of such systems.

\textit{Cancelable biometrics} using multiple traits can make the template more secure and irreversible. The combination of cancelable biometric with biometric encryption techniques is a topic to look forward to research for solving the problem of stolen biometrics. 

\textit{2D-3D biometrics} is the transformation of the 2D biometric image into a 3D image to be used for enrollment and authentication. The image processing techniques that help in such transformation can explore new horizon.

\textit{Multimodal biometric system} using more than one trait to enroll and authorize an individual is gaining more attention by the research community. The performance and security of such system have a vast scope in contributing to biometric system future.

\textit{Biometrics as digital signature} It is desirable to generate a digital signature using biometrics but not practicable because of its inaccurate measuring and potential hill-climbing attacks, without using specific hardware devices that hold signature keys or biometric templates securely \cite{DBLP:conf/iccsa/KwonL04}. 

\section{Conclusion}\label{conclusion}
In this paper, we have done a comprehensive study of the fingerprint biometric system and addressed various security aspects of the system. The security vulnerabilities of a biometric system were highlighted using threat models. The direct, indirect and side channel attacks targeting biometric system were elaborated to understand the weaknesses in the system. The application of cryptographic techniques in securing biometric system termed as biometric cryptosystem and cancelable biometrics has contributed significantly to the template protection techniques. We have identified some of the open problems and challenges that need to be addressed by the researchers. The biometric security has a lot of scopes to make further the system more invulnerable to different types of attacks. We mainly focused on including the topics related to the biometric system that is sufficient enough to build a base for the reader and encourage them to further explore and contribute in the field of biometric security.

\bibliographystyle{IEEEtran}
\bibliography{Survey}

\end{document}